\documentclass[sigconf]{acmart}
\AtBeginDocument{%
  }

\usepackage{balance}
\usepackage{times}
\usepackage{latexsym}
\usepackage[T1]{fontenc}
\usepackage[utf8]{inputenc}
\usepackage{microtype}
\usepackage{bigstrut}
\usepackage{boldline}
\usepackage{graphicx}
\usepackage{filecontents}
\usepackage{amsfonts}
\usepackage{amsmath}
\usepackage{bbm}
\usepackage{booktabs}
\usepackage{xspace}
\usepackage{cleveref}
\usepackage{enumitem}
\usepackage{graphicx}
\usepackage{multirow}
\usepackage{subcaption}
\usepackage{array}
\usepackage{makecell}
\usepackage{tikz}
\usepackage{colortbl}
\usepackage{xcolor}
\usepackage{listings, listings-rust}
\usepackage{soul}
\usepackage{titlesec}

\titleformat{\subsection}
  {\normalfont\fontsize{10}{15}\bfseries}{\thesubsection}{1em}{}

\newcommand\javadocblue[1]{\textcolor[rgb]{0.25,0.35,0.75}{{#1}}}

\definecolor{blueish}{RGB}{250, 250, 255}
\definecolor{greenish}{RGB}{250, 255, 250}
\definecolor{redish}{RGB}{255, 200, 200}

\definecolor{highlight}{RGB}{175, 255, 100}
\definecolor{gray01}{gray}{.98}
\definecolor{gray05}{gray}{0.95}
\definecolor{gray08}{gray}{0.92}
\definecolor{gray10}{gray}{0.90}
\definecolor{gray12}{gray}{0.88}
\definecolor{gray15}{gray}{0.85}
\definecolor{gray18}{gray}{0.82}
\definecolor{gray20}{gray}{0.80}
\definecolor{gray25}{gray}{0.75}
\definecolor{gray30}{gray}{0.70}
\definecolor{gray35}{gray}{0.65}
\definecolor{gray40}{gray}{0.60}
\definecolor{gray45}{gray}{0.55}
\definecolor{gray50}{gray}{0.50}
\definecolor{gray55}{gray}{0.45}
\definecolor{gray60}{gray}{0.40}
\definecolor{gray65}{gray}{0.35}
\definecolor{gray70}{gray}{0.30}
\definecolor{gray75}{gray}{0.25}
\definecolor{gray80}{gray}{0.20}
\definecolor{gray85}{gray}{0.15}
\definecolor{gray90}{gray}{0.10}
\definecolor{gray95}{gray}{0.05}
\usepackage{enumitem}

\newcommand{\bi}{\begin{itemize}[leftmargin=*, wide=0pt, itemsep=0pt, parsep=0pt]}
\newcommand{\ei}{\end{itemize}}
\newcommand{\beq}{\begin{equation}}
\newcommand{\eeq}{\end{equation}}
\newcommand{\be}{\begin{enumerate}[leftmargin=*, wide=0pt, itemsep=0pt, parsep=0pt]}
\newcommand{\ee}{\end{enumerate}}

\newcommand{\sysname}{\textsc{SpecTra}\xspace}

\newcommand\res[1]{}

\newcommand\tion[1]{\hbox{\S\ref{sect:#1}}\xspace}

\newcommand{\eg}{\hbox{\emph{e.g., }}\xspace}
\newcommand{\ie}{\hbox{\emph{i.e., }}\xspace}

\definecolor{javared}{rgb}{0.6,0,0} % for strings
\definecolor{javagreen}{rgb}{0.25,0.5,0.35} % comments
\definecolor{javapurple}{rgb}{0.5,0,0.35} % keywords
\definecolor{javadocblue}{rgb}{0.25,0.35,0.75} % javadoc

\newcommand{\TT}[1]{{\javadocblue{\small\texttt{#1}}}}

\definecolor{bluekeywords}{rgb}{0,0,1}
\definecolor{greencomments}{rgb}{0,0.5,0}
\definecolor{greennumbers}{rgb}{0,0.5,0}
\definecolor{redstrings}{rgb}{0.64,0.08,0.08}
\definecolor{blueident}{rgb}{0, 0.08, 0.45}

\lstdefinestyle{myc}{
  language=c,
  showstringspaces=false,
  basicstyle=\small\ttfamily,
  keywordstyle=\color{bluekeywords},
  commentstyle=\color{greencomments},
  identifierstyle=\color{blueident},
  stringstyle=\color{redstrings},
  frame=none,
  backgroundcolor=\color{white},
  escapeinside={<@}{@>},
  columns=fullflexible,
  breakatwhitespace=true,
  breaklines=true,
  literate=
  *{0}{{{\color{greennumbers}0}}}1
  {1}{{{\color{greennumbers}1}}}1
  {2}{{{\color{greennumbers}2}}}1
  {3}{{{\color{greennumbers}3}}}1
  {4}{{{\color{greennumbers}4}}}1
  {5}{{{\color{greennumbers}5}}}1
  {6}{{{\color{greennumbers}6}}}1
  {7}{{{\color{greennumbers}7}}}1
  {8}{{{\color{greennumbers}8}}}1
  {9}{{{\color{greennumbers}9}}}1
}

\definecolor{yellowhl}{rgb}{1,0.91,0.75}
\sethlcolor{yellowhl}

\newcommand*\circled[1]{\tikz[baseline=(char.base)]{
            \node[shape=circle,fill=gray,inner sep=2pt] (char) {\footnotesize\textcolor{white}{\bfseries #1}};}}

\titlespacing\subsection{0pt}{6pt plus 4pt minus 2pt}{0pt plus 2pt minus 2pt}

\setcopyright{acmlicensed}
\copyrightyear{2018}
\acmYear{2018}
\acmDOI{XXXXXXX.XXXXXXX}
\acmConference[Conference acronym 'XX]{Make sure to enter the correct
  conference title from your rights confirmation email}{June 03--05,
  2018}{Woodstock, NY}
\acmISBN{978-1-4503-XXXX-X/2018/06}

\begin{document}

%\title{\sysname{}: Enhancing the Code Translation Ability of Language Models by Generating Multi-Modal Specifications}
\title{\sysname{}: Enhancing LLM-based Code Translation with Multi-Modal Specifications}

\author{Vikram Nitin}
\email{vikram.nitin@columbia.edu}
\affiliation{%
  \institution{Department of Computer Science,\\Columbia University}
  \city{New York}
  \state{NY}
  \country{USA}
}

\author{Rahul Krishna}
\email{rkrsn@ibm.com}
\affiliation{%
  \institution{IBM Research}
  \city{Yorktown Heights}
  \state{NY}
  \country{USA}}

\author{Baishakhi Ray}
\email{rayb@cs.columbia.edu}
\affiliation{%
  \institution{Department of Computer Science,\\Columbia University}
  \city{New York}
  \state{NY}
  \country{USA}
}
\renewcommand{\shortauthors}{Trovato et al.}

\begin{abstract}
Large language models (LLMs) are increasingly being used for the task of automated code translation, which has important real-world applications. However, most existing approaches use only the source code of a program as an input to an LLM, and do not consider the different kinds of specifications that can be extracted from a program. In this paper, we propose \sysname{}, a multi-stage approach that uses a novel self-consistency filter to first generate high-quality static specifications, test cases, and natural language descriptions from a given program, and then uses these along with the source code to improve the quality of LLM-generated translations. We evaluate \sysname{} on three code translation tasks - C to Rust, C to Go, and JavaScript to TypeScript - and show that it can enhance the performance of six popular LLMs on these tasks by up to a relative improvement of 46\%. We also present a case study on extending this approach to handle translation of a full C project to Rust. Our research suggests that generating high-quality specifications could be a promising and efficient way to improve the performance of LLMs for code translation.
\end{abstract}

\begin{CCSXML}
<ccs2012>
   <concept>
        <concept_id>10011007.10011074.10011099.10011102.10011103</concept_id>
       <concept_desc>Software and its engineering~Software testing and debugging</concept_desc>
       <concept_significance>500</concept_significance>
       </concept>
   <concept>
       <concept_id>10011007.10010940.10010992.10010998.10011000</concept_id>
       <concept_desc>Software and its engineering~Automated static analysis</concept_desc>
       <concept_significance>300</concept_significance>
       </concept>
   <concept>
       <concept_id>10011007.10011006.10011041</concept_id>
       <concept_desc>Software and its engineering~Compilers</concept_desc>
       <concept_significance>300</concept_significance>
       </concept>
   <concept>
       <concept_id>10011007.10011074.10011099</concept_id>
       <concept_desc>Software and its engineering~Software verification and validation</concept_desc>
       <concept_significance>300</concept_significance>
       </concept>
   <concept>
       <concept_id>10002951.10003317.10003338.10003341</concept_id>
       <concept_desc>Information systems~Language models</concept_desc>
       <concept_significance>300</concept_significance>
       </concept>
 </ccs2012>
\end{CCSXML}



\maketitle

\section{Introduction}
\label{sec:intro}

Code translation refers to the process of converting code written in one programming language into functionally equivalent code in another language. This task is particularly important when updating legacy code written in older languages (e.g., C, JavaScript) to modern programming languages (e.g., Rust, Go, TypeScript). Updating legacy code for critical systems become critical as legacy code often incurs high costs due to maintenance overhead, outdated dependencies, and lack of documentation, making debugging and improvements difficult. It also increases security risks, as older systems lack critical updates and are more vulnerable to modern threats. In fact, the ``technical debt" associated with legacy code maintenance costs the United States more than one trillion dollars annually~\cite{mims2024invisible}. Consequently, translating legacy code into modern languages is an urgent priority.

Traditionally, program translation has been achieved through transpilers~\cite{c2rust, feldman1990fortran}. Transpilers are language-specific, rule-based systems utilizing carefully designed algorithms and heuristics to produce code in the target language, where the transpiled programs are semantically equivalent (within compilation framework gurantee) to behave identically to the original source code. Although such translations are functionally equivalent, they frequently lack idiomatic structure—meaning that the resulting code differs significantly from typical human-authored programs. For example, transpiler-generated code may exhibit unusual control flow patterns, rely on uninformative or machine-generated identifier names, invoke foreign functions from the source language, and perpetuate the inherent problems of the original legacy code. Such non-idiomatic code can therefore be challenging to read, maintain, and enhance, ultimately undermining the intended benefits of code translation. To address these issues,  a growing body of research~\cite{roziere2020unsupervised, ahmad2022summarize, roziere2021leveraging, pan2024lost, yang2024exploring} investigates the use of LLMs for automatic translation between programming languages. In contrast to transpilers, LLM-generated translations tend to produce code that is human-readable and maintainable, but provides fewer guarantees regarding correctness (see~\Cref{sec:motivating}).

In this paper, we propose improving LLM-based code translation by integrating core principles from transpiler methodologies. Transpilers rigorously preserve a program's functional specifications—ensuring that translated code is semantically equivalent to the original. We adopt this key insight by explicitly embedding these functional specifications into the prompts used by the LLM during translation. Rather than relying solely on the LLM's implicit understanding, we explicitly define functional constraints, guiding the model to produce code that aligns closely with the original semantics.
This approach, thus, aims to adapt the semantic equivalency checking from transpiler methods (although at a very approximate level) with the readability and idiomatic nature of LLM-generated code, providing translations that are both accurate and maintainable.

Semantics of a program can be expressed in different ways, such as formal specifications (abbreviated as ``specs'') written as logic rules, or natural language (NL) specs. Each helps to ensure correctness and clarity in different ways. While formal specs can define the program’s structure, invariants that the programs must always satisfy, etc., NL specs can explain the program’s overall intent in human-readable form, making it easier to connect the formal rules with human understanding. We hypothesize that, to improve translation accuracy, LLMs benefit from both formal and informal specs, as they provide a deeper understanding of the source program in complementary ways. To this end, we use three different types of specifications:

% \begin{itemize}[leftmargin=*]
%     \item \textit{Static Specifications} describe rules that always apply, such as pre- and post- conditions of a function or type constraints. These help define the program’s static properties, regardless of specific inputs.
    
%     \item \textit{Dynamic Specifications} illustrate how the program behaves with real input-output examples. Unlike static spec, they give concrete demonstrations of expected functionality in different situations.
    
%     \item \textit{Natural Language Descriptions} provide informal yet intuitive explanations of the program’s behavior, helping capture human insights that may not be explicitly stated in the code.
% \end{itemize}

\begin{itemize}[leftmargin=*]
    \item \textit{Static Specifications} capture properties of source programs that universally hold for all valid inputs and outputs, such as preconditions, postconditions, and type constraints. These specifications define global correctness criteria independent of particular execution instances. Such specs are often helpful in verification \& validation.
     
     \item \textit{Dynamic Specifications} illustrate how the program behaves with real input-output examples. Unlike static specs, they give concrete demonstrations of expected functionality in different situations and thus, often help in testing and debugging. 

     \item \textit{Natural Language Descriptions}, although inherently less precise, offer intuitive context and capture informal human insights into a program’s intended functionality. Such descriptions can assist the LLM in understanding the broader semantic goals and idiomatic conventions underlying a given code translation task.
\end{itemize}

By combining these three types of specifications, we provide structured constraints and intuitive guidance to the LLM during translation. This approach not only improves the translated code while preserving the original semantics, as traditional transpilers do, but also maintains readability, clarity, and natural coding style.

However, inferring precise functional specifications for arbitrary source programs across multiple programming languages is inherently difficult. Traditional specification inference tools~\cite{taghdiri2007inferring, ramanathan2007static, kang2016apex}, such as Daikon~\cite{perkins2004efficient}, typically focus on specific types of specifications, depend heavily on language-specific instrumentation, and express results using specialized domain-specific languages (DSLs). As a result, these tools may not generalize well across different languages and can be challenging for LLMs to interpret, particularly if the LLMs have not been trained on these DSLs.

To address this limitation, following the previous ML/LLM-based approaches for inferring program properties~\cite{raychev2015predicting, rahmani2021multi, endres2024can}, we employ a state-of-the-art LLM directly as a multi-modal specification annotator. However, there is no guarantee that the LLM generated specs are actually correct. The key novelty of our approach lies in the {\em rigorous validation of these LLM-generated specifications}, ensuring they accurately reflect the semantics of the original source program. Concretely, our validation methodology is tailored specifically to the type of generated specifications:

\begin{itemize}[leftmargin=*]   
    \item For static and natural language specifications, we introduce a novel self-consistency-based validation strategy, inspired by recent advancements in robustness testing of code models~\cite{min2023beyond}. Given a source code snippet, we prompt an LLM to generate a specification and then use this specification to regenerate the code in the original language, without any additional context. If the regenerated code closely matches the original (both written in the same source language), it suggests that the specification accurately captures the program’s essential functionality and can be reliably used for translation. Conversely, significant discrepancies indicate gaps or inaccuracies in the generated specification, necessitating further refinement or rejection.

    \item 
    Validation for dynamic specifications (I/O) is straightforward: we execute the original program using the LLM-generated inputs and verify whether the resulting outputs match those generated by the model. Any mismatch indicates incorrect or incomplete specifications. When discrepancies arise, we correct output errors by replacing the LLM-generated outputs with the actual execution results. For other execution-related errors, we iteratively regenerate the I/O examples, retrying until consistent, accurate specifications are obtained.
\end{itemize}

This iterative validation strategy ensures that only specifications that faithfully encapsulate the source program's original semantics are subsequently used. After validation, we use these high-quality specifications as explicit constraints in translation prompts to the LLM, guiding it towards generating translations that are both idiomatic and semantically correct. To this end, we introduce \sysname{}, a novel approach for specification guided LLM-based code translation. \sysname{} provides a recipe on how to combine specs of different modalities for faithful program translation. Thus, by rigorously validating specification quality through self-consistency checks and combining the specs in optimal order, our approach significantly enhances the reliability, robustness, and effectiveness of LLM-based cross-language code translation.

We evaluate \sysname{} using two settings. First, using self-contained code contest level programs, we perform a thorough empirical analysis of the performance of \sysname{} and its various design decisions. Here we evaluate on three code translation tasks - converting {C} to {Rust}, {C} to {Go}, and {JavaScript} to {TypeScript}. 
There is a lot of interest within the software engineering community in converting between these pairs of languages \cite{google-2021a, google-2021b}, and thus these tasks have direct real-world applicability. 
We find that \sysname{} is able to enhance the performance of 6 popular LLMs on these tasks by up to {46\%} %(relative improvement) 
compared to a baseline without specifications.%~\res{what is uni-modal baseline?}.

% Next, we employ \sysname{} to translate full projects. Due to resource limitations, we only focus on \textbf{C} to \textbf{Rust} full project translation. We use a benchmark consisting of 24 functions taken from the \TT{cat} utility of GNU Coreutils, which are covered by the provided system-level test cases. We show that we are able to successfully extend our specification generation and validation framework to this setting. Further, \sysname{} improves on a non-specification baseline by up to 10\% (relative improvement). 

To further demonstrate the capabilities of \sysname{}, we conduct a case study on full-project translation. While resource constraints limit our focus to {C} to {Rust} translation, we rigorously evaluate \sysname{} on a benchmark of 24 functions from the \TT{cat} utility in GNU Coreutils, all covered by system-level test cases. Our results affirm that our specification generation and validation framework seamlessly extends to this challenging setting. Moreover, \sysname{}  outperforms a no-specification baseline, achieving up to a {10\%} relative improvement, showcasing its effectiveness in producing high-fidelity translations.

To summarize, our contributions are as follows:
\begin{enumerate}
\item We propose a novel approach based on self-consistency to generate and validate formal and informal specifications of a program. 
\item We demonstrate that strategically combining these specifications enhances program translation quality. To realize this, we introduce \sysname{}, a novel framework that seamlessly integrates this approach.
%\item integrate these validated specifications with the source code and propose \sysname{}, an approach that utilizes multi-modal specifications to improve the quality of LLM-generated code translations.
\item We extensively evaluate \sysname{} on three program translation tasks and six popular open source and proprietary LLMs of various sizes. %The findings of this paper indicate that \sysname{} improves the performance of baseline models by up to 46\%.
\item We further conduct a case study on the full-project translation of a \texttt{coreutils} program\cite{brady2017how}. %and demonstrate that \sysname{} outperforms a baseline without specifications, achieving up to a 10\% improvement.
\end{enumerate}
We make our code and data publicly available\footnote{\url{https://github.com/spectra822/icse26}~\res{rename}}, anonymized for review.
\res{give opensource link}

\begin{table*}
    \centering
    \resizebox{0.9\linewidth}{!}{
    \begin{tabular}{ccc}
    \toprule
      % \hline
    \raisebox{-0.57\totalheight}{
        \begin{minipage}{0.3\textwidth}
\lstset{style=myc}
\lstinputlisting{a.c}
\subcaption{A C program with a complicated conditional, highlighted in \hl{yellow}.}
\label{tab:motivating_a}
        \end{minipage}}
        
        &
        
    \raisebox{0\totalheight}{
        \begin{minipage}{0.3\textwidth}
\lstset{style=myc}
\lstinputlisting{b.c}
\subcaption{A static specification for the same C program, generated by an LLM. This also contains an error, highlighted in \hl{yellow}.}
\label{tab:motivating_b}
        \end{minipage}}
        
        &
        
        \begin{minipage}{0.3\textwidth}
\lstset{style=myc}
\lstinputlisting{c.c}
\subcaption{From top to bottom - correct static, I/O and NL specifications for the original C program.}
\label{tab:motivating_c}
        \end{minipage} \\
        % \hline
        \midrule
        
        \begin{minipage}{0.3\textwidth}
\lstset{language=Rust, frame=none, style=colouredRust, basicstyle=\small\ttfamily, escapeinside={<@}{@>}}
\lstinputlisting{d.c}
\subcaption{A Rust translation of the above C program, missing a check on the length of the buffer (the missing check is highlighted in \hl{yellow}).}
\label{tab:motivating_d}
        \end{minipage}
        
        &
        
        \begin{minipage}{0.3\textwidth}
            \lstset{language=Rust, frame=none, style=colouredRust, basicstyle=\small\ttfamily, escapeinside={<@}{@>}}
\lstinputlisting{e.c}
\subcaption{A Rust translation of the C program from \ref{tab:motivating_a}, augmented with the above incorrect specification from \ref{tab:motivating_b}. It incorrectly checks if the length is \textit{odd} instead of \textit{even} (highlighted in \hl{yellow}).}
\label{tab:motivating_e}
        \end{minipage}
        
        &
        
        \raisebox{0.08\totalheight}{
        \begin{minipage}{0.3\textwidth}
\lstset{language=Rust, frame=none, style=colouredRust, basicstyle=\small\ttfamily, escapeinside={<@}{@>}}
\lstinputlisting{f.c}
\subcaption{A correct translation of the C program aided by the above static specification from \ref{tab:motivating_c}.}
\label{tab:motivating_f}
        \end{minipage}}\\
        \bottomrule
        % \hline
    \end{tabular}%
    }
    \vspace{0.5cm}
    \caption{A motivating example demonstrating how correct multi-modal specifications can guide an LLM to produce accurate translations.}
    \label{tab:motivating}
\end{table*}

\section{Motivating Example}
\label{sec:motivating}

In this section, we introduce an example to illustrate how correct specifications can be beneficial for translating code. This example is taken from CodeNet \cite{puri2021codenet}, and the responses are produced by GPT-4o. \Cref{tab:motivating_a} shows a C program which reads an input buffer and prints either \texttt{\textbf{\color{teal!80}"First"}} or \texttt{\textbf{\color{teal!80}"Second"}} based on a condition (highlighted).

This condition seems relatively complex on first glance, and it can be tricky to reason about, even for a human. On further analysis, we can see that it evaluates to \TT{True} if the first and last character of the input are the same \textit{and} the length of the input is even (suggested by the bitwise \texttt{AND} operation \TT{\&1}.).
If we ask an LLM to translate this C code into Rust, as shown in \Cref{tab:motivating_d}, it is able to generate code that reads the input into a buffer and prints \texttt{\textbf{\color{teal!80}"First"}} or \texttt{\textbf{\color{teal!80}"Second"}}. However, it omits the check for the length of the input being even~\res{where is this check in C or Rust code? What is the significance of the highlighted text in yellow?}.

Prior work \cite{wei2022chain} has established that LLMs can benefit from step-by-step prompting on tasks involving complex reasoning. Accordingly, we could ask an LLM to first reason about the properties of the \TT{main()} function by generating a \textit{static specification}. However, the first candidate specification that the LLM generates, shown in \Cref{tab:motivating_b} still has a flaw. This time, the LLM realizes that there is a condition on the length of the input, but it interchanges odd with even. When we try to use this flawed specification to translate the C program, we get an incorrect Rust program shown in \Cref{tab:motivating_e}.

However, if we sample multiple candidate specifications from the LLM by setting a non-zero temperature, we eventually get one that is correct. This is shown at the top of \Cref{tab:motivating_c}. Now, if we provide this correct specification along with the C code, the LLM is successfully able to translate it to Rust, with the correct condition on input length. The resulting translation is shown in \Cref{tab:motivating_f}. Alternatively, we could use \textit{dynamic} specifications that capture specific inputs and outputs of the program, or \textit{natural language} specifications that describe the function in imprecise terms. These are also shown in \Cref{tab:motivating_c}. From this discussion, we can derive the following insights:

%\smallskip
\noindent
\ul{\textbf{Insight \#1:} Correct specifications provided along with code can help an LLM generate a more accurate translation of the code.} \sysname{} filters out incorrect specifications using a novel self-consistency check, thereby avoiding any potential errors in LLM reasoning. It then uses these self-consistent specifications to improve the quality of LLM translations.

\smallskip
\noindent
\ul{\textbf{Insight \#2:} Specifications of multiple modalities can guide the LLM towards a correct translation.} \sysname{} uses three different modalities of specifications, namely static, dynamic and natural language, all of which work in unison to produce better translations.

% \begin{figure}[t!]
%     \centering
%     \includegraphics[width=0.8\linewidth]{figures/figure1.v2.pdf}
%     \caption{Correct program specifications in the prompt can help LLMs in program translations (top figure), whereas translating with no specifications provided (bottom figure) can cause erroneous translations (as shown by incorrect translation due to a missing conditional).}
%     \label{fig:motivating}
% \end{figure}

\section{Methodology}

\begin{figure*}[t!]
    \centering
    \includegraphics[trim={.2mm 0 0 0.2mm},clip, width=\linewidth]{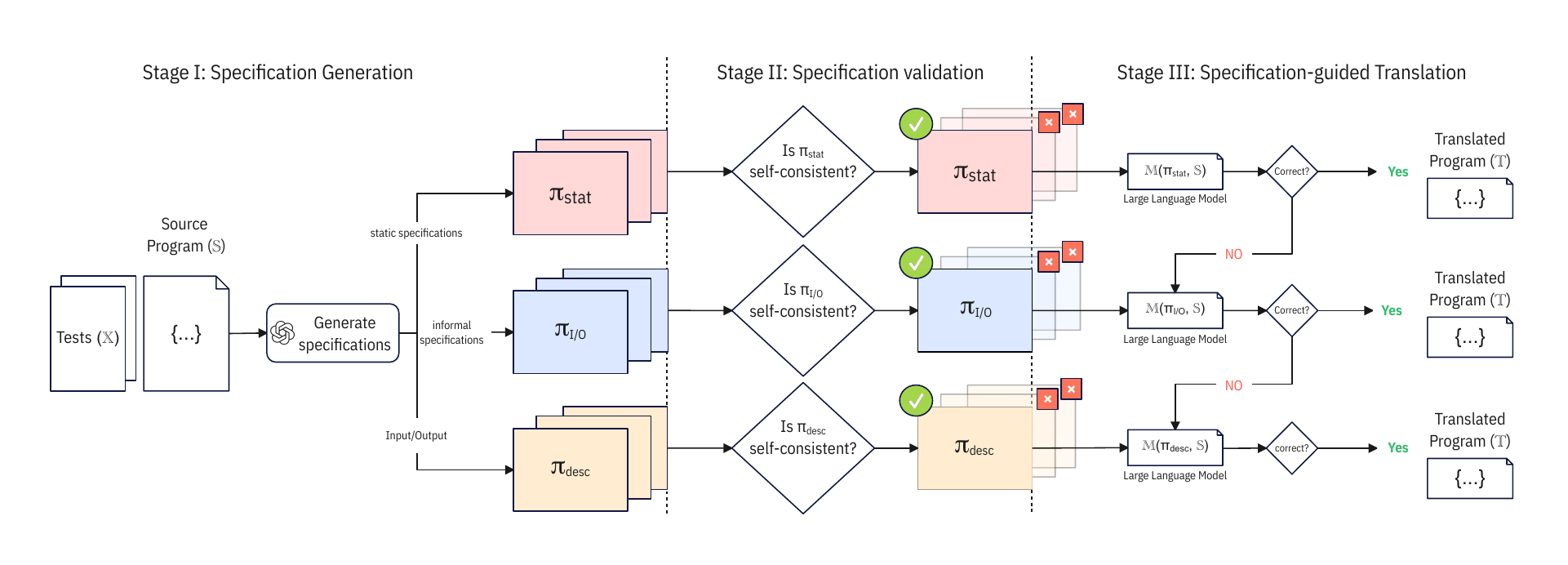}
    \caption{An overview of \sysname{}. }
    \label{fig:overview}
\end{figure*}

% In this section, we describe \sysname{}, a methodology for translating code from one programming language ($\mathcal{S}$) to another language ($\mathcal{T}$) using a large language model ($\mathbb{M}$) by conditioning it on one of many specification modalities ($\mathcal{\pi}$). These modalities encompass static specifications ($\pi_{stat}$), represented as loosely defined sets of program invariants; input-output relations ($\pi_{I/O}$) through test cases; and informal specifications articulated through natural language descriptions ($\pi_{desc}$). Collectively, these approaches aim to enhance the functional correctness of the translated code.
% In this section, we describe \sysname{}, a methodology for translating a program ($\mathcal{S}$) from source language to a target language using a large language model ($\mathbb{M}$) by conditioning it on one of many specification modalities ($\pi$). The validity of the translated program is evaluated by an evaluator ($\mathcal{E}$), which is often a ground-truth test case that run the source and target versions of the progarm with same input and expects same output. 

In this section, we introduce \sysname{}, a methodology for translating a program ($\mathcal{S}$) from a source language to a target language using a large language model ($\mathbb{M}$). This process involves conditioning the model on one of several specification modalities ($\pi$). The validity of the translated program is assessed by an evaluator ($\mathit{eval}$), which typically consists of a ground-truth test case. This test case runs both the source and target versions of the program with the same input and expects identical output.

The methodology consists of three steps: (i) specification generation, (ii) specification validation, and (iii) specification combination for translation. This section details each step in depth.

\subsection{Specification Generation}
\label{sect:stage-1}
% In this work, a language model (GPT4o) generates three types of specifications: 1. static (\tion{}), 2. input-output (\tion{}), and 3. descriptions (\tion{}). For each type of specification, multiple candidates are produced, and the candidate consistent with the input program is selected (discussed in \tion{self-consistency}). Utilizing a large language model of the such as GPT4o ensures correctness and completeness, which are critical for generating reliable specifications. This choice is motivated by an establish precedence in literature for annotating text. 

The first stage of \sysname{} focuses on the generating three types of specifications using a large language model (here, GPT4o): a) static (\tion{spec-static}), b) input-output (\tion{spec-io}), and c) descriptions (\tion{spec-desc}). 
% For each specification type, multiple candidates are generated, and the one consistent with the input program is selected (the validation and selection criteria are discussed in \tion{stage-2}). 
% This approach is supported by established precedents in the literature for annotating code \cite{ahmad2022summarize}.

\subsubsection{Static Specifications} 
\label{sect:spec-static}
Static specifications are a structured representation of the behaviour of section of the program. We use an LLM ($\mathbb{M}'$) and a prompt composing function ($\lambda_\text{stat}$) to take a given program ($\mathcal{S}$) and generate $k$ (here, $k=3$) candidate static specifications ($\pi_{stat}^{i}$) for the given program. This can be formulated as:
\begin{equation}
\label{eq:eq-stat}
\small
\Pi_{stat} = \mathbb{M}'~(~\lambda_\text{stat}(~\mathcal{S}))=\left\{\pi_{stat}^{1}, \pi_{stat}^{2}, \ldots, \pi_{stat}^{k}\right\} 
\end{equation}
Our motivating example, \Cref{tab:motivating}, shows a static specification. The generated static specifications have the following components:

\bi
\item[$\mathcal{\Pi}$] {\textbf{Input Format:}} The initial user input required for the program to execute, \eg {\em a single string of characters without any spaces}.
% \item[$\mathcal{\Pi}$] \textit{Input Format:} The initial conditions or data required for the program to execute, \eg {\em the input provided to the program will be a string ending with a newline character}.
\item[$\mathcal{\Pi}$] {\textbf{Output Format:}} The result or data produced by the program after execution, \eg {\em a single word, either \texttt{\textbf{\color{teal!80}"First"}} or \texttt{\textbf{\color{teal!80}"Second"}}}.
\item[$\mathcal{\Pi}$] {\textbf{Pre-condition:}} The conditions that must be true before the function is executed. \eg {\em the buffer \texttt{\textbf{\color{blue!70}buf}} must be declared and large enough to store the input string.}
% which in this case is \texttt{\textbf{\color{blue!80}char buf[114514];}}.
\item[$\mathcal{\Pi}$] {\textbf{Post-condition:}} The conditions that must be true after the function has executed. \eg {\em Prints \texttt{\textbf{\color{teal!80}"First"}}} if the first and last
characters of the input are the same and the length
of input is even, otherwise prints \texttt{\textbf{\color{teal!80}"Second"}}.
% \item[$\mathcal{\Pi}$] \textit{Post-condition:} The conditions that must be true after the program section has executed. \eg {\em for the input \texttt{\textbf{\color{teal!80}"hello\n"}, since the first and last characters ('h' and 'o') are not the same, the program prints \texttt{\textbf{\color{teal!80}"Second\n"}}.
\ei

%% Moving this to the next stage
% $\lambda_\text{stat-gen}$ takes a program and test case, and creates a text prompt containing instructions to generate a static specification. In the next step, we use the same LLM to re-generate a program from each of the candidate specifications.
% \begin{align*}
% \mathcal{S}'_1, ...,\mathcal{S}'_k =~&\mathbb{M}~(~\lambda_\text{rev-stat-gen}~(\mathcal{I}_1)~)~,\\
% &...,~\mathbb{M}~(~\lambda_\text{rev-stat-gen}~(\mathcal{I}_k)~)
% \end{align*}
% Here $\lambda_\text{rev-stat-gen}$ takes an static specification and creates a text prompt containing instructions to re-generate a source program. Finally, we pick a set $\Pi_{stat}$ of static specifications corresponding to re-generated programs that pass the test case.
% \[
% \Pi_{stat} = \{\mathcal{I}_i~:~\mathcal{S}'_i(X) = Y\}_{i=0}^k
% \]

\subsubsection{Input-output Specifications} 
\label{sect:spec-io}

% \begin{figure}[t]
% \centering
% \lstset{style=myc, basicstyle=\footnotesize\ttfamily}
% \begin{lstlisting}
% void convert(char charNum[], int intNum[])
% // Precondition: `charNum` is a null-terminated string representing a non-negative integer, with each character being a digit ('0'-'9').
% // Postcondition: The integer representation of `charNum` has been stored in reverse order into the array `intNum`.

% int main(int argc, const char * argv[])
% // Input format: The first line contains an integer `dataSet`. For each dataset, two lines are provided, each containing a string representing a non-negative integer.
% // Output format: For each dataset, a single line is printed...
% // Precondition: The input integers are provided as strings, and their lengths do not exceed 128 characters.
% // Postcondition: The program calculates the sum of two input integers for each dataset, checks for overflow, and prints the result or "overflow" as specified.
% \end{lstlisting}
% \caption{A static specification for a C program with multiple functions. Our static specifications and descriptions are generated at the function level.}
% \label{fig:multifunc-example}
% \end{figure}

Input-output specifications represent the expected behavior (\ie output) of a section of a program given a specific input. As with the static specification, $k$ input-output specifications ($\pi_{I/O}^{k}$) are generated using an LLM $\mathbb{M}'$ for the given program.
% \[
% (X'_1, Y'_1),~...~,~(X'_k, Y'_k) = \mathbb{M}~(~\lambda_\text{test-gen}(~\mathcal{S}~)~)
% \]
We make use of another prompt composing function ($\lambda_\text{I/O}$) to take a program and create a text prompt containing instructions to generate test inputs (${x\prime}$) and corresponding outputs (${y^\prime}$). In other words, $\pi_{I/O}^{i} = (x_i^\prime, y_i^\prime)$.
% \begin{equation}
% \label{sect:io-spec-pairs}
% \small
% \pi_{I/O}^{i}=\{x_{i}^{\prime}, y_{i}^{\prime}\}\quad\forall i\in\{1,...,k\}
% \end{equation}
In summary, the input-output specification generation process may be formulated as:
\begin{equation}
\label{eq:eq-io}
\small
\Pi_{I/O} = \mathbb{M}'~(~\lambda_{I/O}(~\mathcal{S}))=\left\{\pi_{I/O}^{1}, \pi_{I/O}^{2}, \ldots, \pi_{I/O}^{k}\right\} 
\end{equation}

% Next, we pick the set $\Pi_{I/O}$ of tests that are consistent with the source program.
% \[
% \Pi_{I/O} = \{(X'_i, Y'_i)~:~\mathcal{S}(X'_i) = Y'_i \}_{i=0}^k
% \]
% Here, $\mathcal{S}(X'_i)$ denotes the output of running $\mathcal{S}$ with $X'_i$ as input. In practice, if we are unable to generate any consistent tests through this approach, \ie if $\Pi_{I/O} = \phi$, then we augment it by pairing each generated input with the \textit{actual} program output, provided that the program exits normally and doesn't raise an error.
% \[
% \Pi_{I/O}~:=~\Pi_{I/O}~\cup~\{~(X'_i, \mathcal{S}(X'_i))~\}_{i=0}^k
% \]

\subsubsection{Generating Descriptions} 
\label{sect:spec-desc}

Program descriptions are an informal and a free-form textual summary of the code to be translated. The procedure to generate these descriptions follows a similar process to that of generating static specifications (as described in \tion{spec-static}). The distinction lies within the prompt-composing function $\lambda_\text{desc}$ which takes the given program ($\mathcal{S}$) and instructs an LLM $\mathbb{M}'$ to generate $k$ the descriptions $\Pi_{desc}=\left\{\pi_{desc}^{1}, \pi_{desc}^{2}, \ldots, \pi_{desc}^{k}\right\}$. This may be represented mathematically as follows:
\begin{equation}
\label{eq:eq-desc}
\small
\Pi_{desc} = \mathbb{M}'~(~\lambda_{desc}(~\mathcal{S}))=\left\{\pi_{desc}^{1}, \pi_{desc}^{2}, \ldots, \pi_{desc}^{k}\right\} 
\end{equation}

% \begin{align*}
% \mathcal{D}_1, ..., \mathcal{D}_k~&= \mathbb{M}~(~\pi_\text{desc-gen}(~\mathcal{P}, (X, Y)~)~)\\
% \mathcal{S}'_1, ...,\mathcal{S}'_k~&= \mathbb{M}~(~\lambda_\text{rev-desc-gen}~(\mathcal{D}_1)~)~,\\
% &~~~...~,~\mathbb{M}~(~\lambda_\text{rev-desc-gen}~(\mathcal{D}_k)~)\\
% \Pi_{desc} &= \{\mathcal{D}_i~:~\mathcal{S}'_i(X) = Y\}_{i=0}^k
% \end{align*}

\subsection{Specification validation}
\label{sect:stage-2}

Specifications generated in the previous stage may be incomplete and/or incorrect. In order to best assist a language model in generating correct code translations, it would be most beneficial to provide complete and verifiable specification. Such specifications are termed as being \textit{self-consistent}.
The objective of this stage, is to retain only \textit{self-consistent} specifications from the candidate specifications.

For the set of static specifications ($\Pi_{stat}$) and descriptions ($\Pi_{desc}$), we use the same language model ($\mathbb{M}'$) to regenerate the \textit{original} source code ($\mathcal{S}$) with a prompt-composition function ($\lambda_{codegen}$). This produces a variant of the source code ($\mathcal{S}_{stat|desc}^{i}$). That is, 

\begin{equation}
\label{eq:eq-desc}
\footnotesize
\mathcal{S}_{stat|desc}^{i} = \mathbb{M}'~\left(~\lambda_{codegen}(\pi_{stat|desc}^{i})\right)\forall i \in \{1, ..., k\} 
\end{equation}

\res{This para is not at paragraph boundary}

\noindent
We define an evaluator $\mathit{eval}(\cdot)$ to compare the equivalence of the variants of the source code $\mathcal{S}_{stat|desc}^{i}$ that were generated using $\pi_{stat|desc}^{i}$ with original source code $S$. \res{What is $\epsilon$} We may represent this as follows:

\begin{equation}
\label{eq:eq-eval}
\footnotesize
\mathit{eval}(\mathcal{S}_{stat|desc}^{i}, \mathcal{S}) = 
\begin{cases} 
\text{True} & \text{if } \mathcal{S}_{stat|desc}^{i}\equiv\mathcal{S} \\
\text{False} & \mathit{otherwise}
\end{cases}
\end{equation}

Our design allows for a variety of comparator functions to be employed. These may be heuristics that measure code similarity (such as BM25 \cite{robertson1994some}), or distinct test cases that exercise the program in a specific manner and expect a pre-determined output. This work employs the latter approach in that we used existing test input and expected output pairs (denoted by $\left(\mathrm{X}, \mathrm{Y}\right)=\{(x_1, y_1), (x_2, y_2), \dots, (x_n, y_n)\}$) as evaluators. Specifically, our evaluation function would assess if the regenerated code $\mathcal{S}_{stat}^{i}$ behaves the same as the original source code $\mathcal{S}$ for all tests.

All the static specifications and descriptions for which the above evaluation fails are discarded.
% Assuming $\epsilon(\mathrm{X}, \mathrm{Y})$ denotes an evaluator that uses the outcome of the tests $(X, Y)$, this selection criteria may be formally described as:
\begin{equation}
\label{eq:eq-select}
\small
\Pi_{stat|desc} = \{\pi_{stat|desc}^{i} \mid \mathit{eval}(\mathcal{S}_{stat|desc}^{i}, \mathcal{S}) = \mathit{True}\}
\end{equation}

For the input-output specifications ($\Pi_{I/O}$), we exercise the input program $\mathcal{S}$ with each $\pi_{I/O}^{i}=\{x_{i}^{\prime}, y_{i}^{\prime}\}$ and discard ones that are inconsistent.
% \begin{equation}
% \label{eq:eq-select}
% \small
% \Pi_{I/O} = \{\pi_{I/O}^{i} \mid \mathit{eval}({S}_{\pi_{I/O}^{i}}, \epsilon(\mathrm{X}, \mathrm{Y})) = \mathit{True}\}
% \end{equation}
% Note, in exceptional cases where discarding inconsistent $\pi_{I/O}$ results in an empty specification set ($\Pi_{i/o}=\emptyset$), we transform the specification to keep the generated input $x_{i}^{\prime}$ and the actual program outputs corresponding to these generated input. 
\begin{equation}
\label{eq:eq-select}
\small
\Pi_{I/O} = \{\pi_{I/O}^{i} \mid S(x'_i) = y'_i\}
\end{equation}
Here, $S(x'_i)$ denotes the output of running the program $S$ using $x'_i$ as input. Note, in cases where discarding inconsistent $\pi_{I/O}$ results in an empty specification set ($\Pi_{i/o}=\emptyset$), we transform the specification to keep the generated input $x_{i}^{\prime}$ and the \textit{actual} program outputs $S(x'_i)$.

% That is,
% \begin{equation*}
% \small
% \Pi_{i/o}=\emptyset \implies \Pi_{I/O} = \pi_{I/O}^{i} 
% \end{equation*}

% The above source code variants are then validated against test input and expected output pairs $\{(x_1, y_1), (x_2, y_2), \dots, (x_n, y_n)\}$ that accompany the source code.

% It is worth noting that, while generating static specifications and descriptions, we take the help of the provided test cases; however our approach can also be extended to a setting where test cases are not provided, by using the test cases that  we generated in the first step, in place of the provided tests. We defer this analysis to future work.

\noindent
\textbf{Handling Multi-Function Standalone Programs.} Our static specifications and descriptions are generated \textit{per function}, \ie we ask the specification-generation model to annotate each function with a specification. This design choice allows our technique to scale to programs with many functions, where a single specification may not sufficiently express the behavior of the entire program. 
% \Cref{fig:multifunc-example} shows a static specification for a C program containing 2 functions.
While verifying self-consistency, we start with a multi-function specification and ask an LLM to generate the \textit{entire} program.

For I/O specifications, however, we generate only a single specification for an \textit{entire} program. It is difficult to describe and validate I/O specifications for individual functions, because the arguments and return types can be arbitrarily complex types like nested structures or pointers. Generating program-level test cases sidesteps this difficulty. Further, since these programs are relatively short, the LLM can potentially reason about the flow of information through the program, using the program-level test cases. However, this assumption is no longer valid when we consider large software projects. This is discussed in more detail in \Cref{sect:full-project}.

\subsection{Specification-guided Translation}
\label{sect:stage-3}

Having generated and validated the specifications $\pi_{stat}$, $\pi_{I/O}$, and $\pi_{desc}$, we now pick one specification at random for each type (say) $\pi_{stat}^{i}$, $\pi_{I/O}^{i}$, and $\pi_{desc}^{i}$ from $\Pi_{stat}$, $\Pi_{I/O}$, and $\Pi_{desc}$ respectively. Next we use these specifications sequentially one at a time along with $\mathcal{S}$ to generate a prompt with a prompt-composer $\lambda_{spec}\left(\mathcal{S}, \pi^{i}\right)$. This prompt is then used to instruct a language model $\mathbb{M}$ to generate translations ($\mathcal{T}$) such that: $\mathcal{T}=\mathbb{M}\left(\lambda_{spec}\left(\mathcal{S}, \pi^{i}\right)\right)$.

In this work, we use an ordered sequence of translations starting with the static specification ($\pi_{stat}$). If this approach fails, we then use the input-output specifications ($\pi_{I/O}$), and, if needed, the natural language descriptions ($\pi_{desc}$). The sequence of specifications is determined by their formality and completeness, starting with the most structured ($\pi_{stat}$) and ending with the least formal ($\pi_{desc}$). In the rare case that all of these fail, we fall back on the ``vanilla'' translation with no specifications. A naive approach is to use all the specification modalities together in the same prompt. We avoided this in favor of a multi-stage approach, guided by the following intuition. As we add each successive specification modality, the information that is gained saturates, but the size of the input keeps increasing. It is known \cite{liu2024lost} that LLMs struggle to process information in long contexts, and this could cause the translation accuracy to degrade. 
\res{Why didn't you give all together?}

\section{Experimental Setup}

% \begin{figure}
%     \centering
%     \includegraphics[width=0.5\textwidth]{figures/boxplot.pdf}
%     \caption{A box plot of the number of attempts needed by \texttt{gpt-4o} to generate a successful specification, for the three types of specifications.}
%     \label{fig:boxplot}
% \end{figure}

\subsection{Datasets}

%\res{talk about two kinds of datasets: one code-context level programs, and full project dataset}

\textbf{Code-Contest Dataset.}
We evaluate our approach on a subset of the CodeNet dataset \cite{puri2021codenet}. CodeNet contains 4053 competitive coding problems, and over 13 million code submissions in a variety of languages including C and JavaScript.
% These submissions are licensed for public use and redistribution, and anonymized to protect the identity of the authors.
We first filter out the problems which don't have test cases. At random, we select 300 C solutions and 300 JavaScript solutions from the remaining problems, making sure that we don't pick more than one solution for the same problem. These 300 C and 300 JavaScript programs, each with accompanying test cases, comprise our evaluation dataset.

\smallskip

\noindent
\textbf{Full-project Translation Case Study.} 
% The \TT{cat} utility from GNU Coreutils
%For our case study of full-project translation, 
We use the \TT{cat} utility from GNU Coreutils\footnote{\url{https://github.com/coreutils/coreutils}}. It consists of over a hundred functions, but several of these are in header files that are shared among multiple coreutils utilities. We pick a subset of 24 of these functions, that are covered by the end-to-end test cases provided for \TT{cat} in the Coreutils repository. %These comprise our dataset for full project translation. 
They have 1452 lines of code, with an average of 60.5 lines per function. The longest function is 522 lines long, and the shortest function is 5 lines long.

%\subsection{Implementation}
% Here, we discuss some of the details of the implementation of full-project translation. Since this is a limited case study, we use a single language pair: C to Rust.
We write custom LLVM/Clang \cite{lattner2004llvm} compiler passes to decompose the entire C program and extract the source code and docstrings of individual C functions. We also instrument the code by adding an LLVM pass that inserts \TT{fprintf} statements at the end of each function body, just before the return statement, to log the arguments and returned value. Printing structures and pointers needs special care. We iterate through individual fields of structures and log their values. For pointers, if it is a \TT{char*} pointer, then we treat it as a string and print up to its first 100 characters. If it is any other pointer type, then we simply print its address.

Once we translate a function from C to Rust, we need to make sure that the code runs end-to-end. To do this, we use Rust's C FFI (Foreign Function Interface) to call functions in C from Rust and vice-versa. We use bindgen\footnote{\url{https://github.com/rust-lang/rust-bindgen}} to generate bindings for each function in C. For more details, refer to our code implementation.

% When we translate a function from C to Rust, its signature in Rust needs to be identical to its signature in C, with C types replaced with Rust proxies for C types. For instance, \TT{char*} in C must be replaced with \TT{*mut libc::c\_char} in Rust. If the signature is not compatible with these requirements, the function will become inaccesible from the C code and could break dependencies. However, this raises another problem - the Rust proxies for C types are \textit{non-idiomatic} and often necessitate the use of \textit{unsafe} Rust code \cite{emre2021translating}. This defeats the primary purpose of translating code from C to Rust, which is to benefit from the enhanced memory and thread safety guarantees of Rust.

% We mitigate this difficulty by generating \textit{two} Rust functions for each C function - and an idiomatic Rust translated function, and a \textit{wrapper function} that calls the idiomatic function. The wrapper function has the signature required to maintain compatibility with the C code, and it handles conversion between non-idiomatic and idiomatic types.

\subsection{Models}

We evaluate on six large language models: (a)~4 proprietary LLMs: \texttt{gpt-4o} and \texttt{gpt-4o-mini} from OpenAI, \texttt{claude-3-sonnet-v2} from Anthropic; \texttt{gemini-1.5-pro} from Google; and (b)~2 open-source LLMs: Mistral Large 2 (123B), and Llama3.1 (70B). We accessed these models through their OpenAI or VertexAI APIs. For the case study on full-project translation, we use only GPT-4o.
% For all these models, we run our experiments against a fixed checkpoint so that our results don't vary with   `q1w2edfnew model releases.
To generate multiple candidate specifications, we use a temperature of 0.6. To generate a single candidate translation, we use greedy decoding with temperature 0, and if we need multiple candidate translations, we use a temperature of 0.3.

\subsection{Evaluation}
\label{sect:eval}

To evaluate a translated program, we compile it, and run the executable with the provided ground truth test input. If the output matches the ground truth  test output, then we mark the translation as correct. If the program doesn't compile, exits with an error code, or if the output doesn't match, then we mark this as incorrect.~\res{this para is not clear. For what you are running output and validating against GT output?}

For running \sysname{} on a program, we generate a list of specifications of different modalities in this order - $\left[\pi_{stat},\pi_{I/O}, \pi_{desc}, \text{none}\right]$. If we were unable to generate one modality of specification, then we omit it from the list. We generate 3 candidate translations, using 3 steps. In step 1, we generate translations with just the \textit{first} specification from the above list and check whether the translation is correct. Next, for step 2, we generate a pair of translations using the first \textit{two} specifications from the list, and check whether \textit{either} of them is correct. Lastly, for step 3, we translate using the first \textit{three} specifications from the list and check whether \textit{any} of them is correct. Then, pass@1, pass@2 and pass@3 for \sysname{} correspond to the accuracy in step 1, step 2 and step 3 respectively. In order to establish a baseline, we generate three translations \textit{without} specifications and similarly measure pass@1 as the accuracy of just the first translation, pass@2 as the accuracy of the first \textit{or} the second, and so on.

%\res{List down the RQs}

\noindent
To this end, we seek to answer the following research questions, evaluating various design choices of \sysname{} on smaller standalone programs (see~\Cref{sec:results}).
\bi
\item \textbf{RQ1: \sysname{} Performance with Multi-modal Specs.} What is the overall performance of \sysname{}?
\item \textbf{RQ2: Impact of Individual Specification Modalities.} How does each individual specification modality guide the LLM towards better translations?
\item \textbf{RQ3: Design Choices.} How do the different components of our design, such as the multi-stage pipeline and the self-consistency check, contribute to the overall performance?
%How can we use all three modalities together to produce more accurate translations?
\item \textbf{RQ4: Evaluating the Generated Specifications.} What is the quality of the generated specifications?
\ei

Once we know the right recipe to configure \sysname{},  we perform a case study to evaluate whether \sysname{} can be effectively extend to translate full software projects, as discussed in Section~\ref{sec:results_fp}.
\section{Empirical Results}%: Translating Smaller Standalone Programs}
\label{sec:results}

Here, we measure \sysname{} on smaller standalone programs. This helps us to find out the optimal configuration of \sysname{} that can be applied to larger real-world setting. 

\begin{table*}[htp!]
\caption{Comparing \sysname{} to the baseline for C to Rust and C to Go translations. The left half of each table shows the absolute number of correct translations out of 300 problems. The absolute number for \sysname{} is highlighted in \colorbox{orange!10}{orange}. Cells highlighted in \colorbox{blue!10}{blue} show higher percentage of correct translations compared to the baseline. We see that \sysname{} outperforms the baseline in all cases.
% \todo{Please add few lines of summary of the colors and findings. Please see Table 4 and use the same about of descriptiveness and detail}
}
\label{tab:c2rust-and-c2go}

\resizebox{0.45\linewidth}{!}{
\begin{tabular}{llrrrrrr}
\multicolumn{1}{c}{}   &     \multicolumn{1}{c}{}     &  \multicolumn{6}{c}{\cellcolor{gray!10}{\textsc{\textbf{C to Rust}}}}        \bigstrut\\\clineB{3-8}{2}
 & \multicolumn{1}{lV{2}}{}    & \multicolumn{3}{c|}{No. correct}  & \multicolumn{3}{c|}{improvement \%}  \bigstrut\\ \clineB{3-8}{2}
& \multicolumn{1}{lV{2}}{}     & \multicolumn{1}{lV{2}}{\rotatebox{90}{pass@1~}}   & \multicolumn{1}{lV{2}}{\rotatebox{90}{pass@2~}}     & \multicolumn{1}{lV{2}}{\rotatebox{90}{pass@3~}}     & \multicolumn{1}{lV{2}}{\rotatebox{90}{pass@1~}}    & \multicolumn{1}{lV{2}}{\rotatebox{90}{pass@2~}}  & \multicolumn{1}{lV{2}}{\rotatebox{90}{pass@3~}}    \bigstrut\\ \clineB{3-8}{2}
&   & \multicolumn{1}{l}{}     & \multicolumn{1}{l}{}     & \multicolumn{1}{l}{}     & \multicolumn{1}{l}{}     & \multicolumn{1}{l}{}     & \multicolumn{1}{l}{}         \bigstrut\\[-1.2em] \hlineB{2}
\multicolumn{1}{V{2}lV{2}}{} &
 \multicolumn{1}{lV{2}}{\textsc{\textsc{Baseline}}} &
  \multicolumn{1}{rV{2}}{176} &
  \multicolumn{1}{rV{2}}{192} &
  \multicolumn{1}{rV{2}}{202} &
  \multicolumn{1}{rV{2}}{$\cdot$} &
  \multicolumn{1}{rV{2}}{$\cdot$} &
  \multicolumn{1}{rV{2}}{$\cdot$} \bigstrut\\ \cline{2-2}
\multicolumn{1}{V{2}lV{2}}{\multirow{-2}{*}{Gpt4o}} &
  \multicolumn{1}{lV{2}}{\sysname{}} &
  \multicolumn{1}{rV{2}}{\cellcolor{orange!10}196} &
  \multicolumn{1}{rV{2}}{\cellcolor{orange!10}219} &
  \multicolumn{1}{rV{2}}{\cellcolor{orange!10}220} &
  \multicolumn{1}{rV{2}}{\cellcolor{blue!10}11\%} &
  \multicolumn{1}{rV{2}}{\cellcolor{blue!10}14\%} &
  \multicolumn{1}{rV{2}}{\cellcolor{blue!10}9\%}\bigstrut\\ \hline
\multicolumn{1}{V{2}lV{2}}{} &
 \multicolumn{1}{lV{2}}{\textsc{\textsc{Baseline}}} &
  \multicolumn{1}{rV{2}}{112} &
  \multicolumn{1}{rV{2}}{125} &
  \multicolumn{1}{rV{2}}{134} &
  \multicolumn{1}{rV{2}}{$\cdot$} &
  \multicolumn{1}{rV{2}}{$\cdot$} &
  \multicolumn{1}{rV{2}}{$\cdot$} \bigstrut\\ \cline{2-2}
\multicolumn{1}{V{2}lV{2}}{\multirow{-2}{*}{Gpt4o-mini}} &
  \multicolumn{1}{lV{2}}{\sysname{}} &
  \multicolumn{1}{rV{2}}{\cellcolor{orange!10}124} &
  \multicolumn{1}{rV{2}}{\cellcolor{orange!10}150} &
  \multicolumn{1}{rV{2}}{\cellcolor{orange!10}156} &
  \multicolumn{1}{rV{2}}{\cellcolor{blue!10}11\%} &
  \multicolumn{1}{rV{2}}{\cellcolor{blue!10}20\%} &
  \multicolumn{1}{rV{2}}{\cellcolor{blue!10}16\%}\bigstrut\\ \hline
\multicolumn{1}{V{2}lV{2}}{} &
 \multicolumn{1}{lV{2}}{\textsc{\textsc{Baseline}}} &
  \multicolumn{1}{rV{2}}{180} &
  \multicolumn{1}{rV{2}}{196} &
  \multicolumn{1}{rV{2}}{201} &
  \multicolumn{1}{rV{2}}{$\cdot$} &
  \multicolumn{1}{rV{2}}{$\cdot$} &
  \multicolumn{1}{rV{2}}{$\cdot$} \bigstrut\\ \cline{2-2}
\multicolumn{1}{V{2}lV{2}}{\multirow{-2}{*}{Claude}} &
  \multicolumn{1}{lV{2}}{\sysname{}} &
  \multicolumn{1}{rV{2}}{\cellcolor{orange!10}208} &
  \multicolumn{1}{rV{2}}{\cellcolor{orange!10}226} &
  \multicolumn{1}{rV{2}}{\cellcolor{orange!10}227} &
  \multicolumn{1}{rV{2}}{\cellcolor{blue!10}16\%} &
  \multicolumn{1}{rV{2}}{\cellcolor{blue!10}15\%} &
  \multicolumn{1}{rV{2}}{\cellcolor{blue!10}13\%}\bigstrut\\ \hline
\multicolumn{1}{V{2}lV{2}}{} &
 \multicolumn{1}{lV{2}}{\textsc{\textsc{Baseline}}} &
  \multicolumn{1}{rV{2}}{160} &
  \multicolumn{1}{rV{2}}{174} &
  \multicolumn{1}{rV{2}}{181} &
  \multicolumn{1}{rV{2}}{$\cdot$} &
  \multicolumn{1}{rV{2}}{$\cdot$} &
  \multicolumn{1}{rV{2}}{$\cdot$} \bigstrut\\ \cline{2-2}
\multicolumn{1}{V{2}lV{2}}{\multirow{-2}{*}{Gemini}} &
  \multicolumn{1}{lV{2}}{\sysname{}} &
  \multicolumn{1}{rV{2}}{\cellcolor{orange!10}171} &
  \multicolumn{1}{rV{2}}{\cellcolor{orange!10}194} &
  \multicolumn{1}{rV{2}}{\cellcolor{orange!10}199} &
  \multicolumn{1}{rV{2}}{\cellcolor{blue!10}7\%} &
  \multicolumn{1}{rV{2}}{\cellcolor{blue!10}12\%} &
  \multicolumn{1}{rV{2}}{\cellcolor{blue!10}10\%}\bigstrut\\ \hline
\multicolumn{1}{V{2}lV{2}}{} &
 \multicolumn{1}{lV{2}}{\textsc{\textsc{Baseline}}} &
  \multicolumn{1}{rV{2}}{95} &
  \multicolumn{1}{rV{2}}{106} &
  \multicolumn{1}{rV{2}}{117} &
  \multicolumn{1}{rV{2}}{$\cdot$} &
  \multicolumn{1}{rV{2}}{$\cdot$} &
  \multicolumn{1}{rV{2}}{$\cdot$} \bigstrut\\ \cline{2-2}
\multicolumn{1}{V{2}lV{2}}{\multirow{-2}{*}{Llama}} &
  \multicolumn{1}{lV{2}}{\sysname{}} &
  \multicolumn{1}{rV{2}}{\cellcolor{orange!10}106} &
  \multicolumn{1}{rV{2}}{\cellcolor{orange!10}130} &
  \multicolumn{1}{rV{2}}{\cellcolor{orange!10}135} &
  \multicolumn{1}{rV{2}}{\cellcolor{blue!10}12\%} &
  \multicolumn{1}{rV{2}}{\cellcolor{blue!10}23\%} &
  \multicolumn{1}{rV{2}}{\cellcolor{blue!10}15\%}\bigstrut\\ \hline
\multicolumn{1}{V{2}lV{2}}{} &
 \multicolumn{1}{lV{2}}{\textsc{\textsc{Baseline}}} &
  \multicolumn{1}{rV{2}}{101} &
  \multicolumn{1}{rV{2}}{117} &
  \multicolumn{1}{rV{2}}{124} &
  \multicolumn{1}{rV{2}}{$\cdot$} &
  \multicolumn{1}{rV{2}}{$\cdot$} &
  \multicolumn{1}{rV{2}}{$\cdot$} \bigstrut\\ \cline{2-2}
\multicolumn{1}{V{2}lV{2}}{\multirow{-2}{*}{Mistral}} &
  \multicolumn{1}{lV{2}}{\sysname{}} &
  \multicolumn{1}{rV{2}}{\cellcolor{orange!10}147} &
  \multicolumn{1}{rV{2}}{\cellcolor{orange!10}159} &
  \multicolumn{1}{rV{2}}{\cellcolor{orange!10}162} &
  \multicolumn{1}{rV{2}}{\cellcolor{blue!10}46\%} &
  \multicolumn{1}{rV{2}}{\cellcolor{blue!10}36\%} &
  \multicolumn{1}{rV{2}}{\cellcolor{blue!10}31\%}\bigstrut\\ \hlineB{2}
\end{tabular}}
\resizebox{0.45\linewidth}{!}{
\begin{tabular}{llrrrrrr}
\multicolumn{1}{c}{}   &     \multicolumn{1}{c}{}     &  \multicolumn{6}{c}{\cellcolor{gray!10}{\textsc{\textbf{C to Go}}}}        \bigstrut\\\clineB{3-8}{2}
 & \multicolumn{1}{lV{2}}{}    & \multicolumn{3}{c|}{No. correct}  & \multicolumn{3}{c|}{improvement \%}  \bigstrut\\ \clineB{3-8}{2}
& \multicolumn{1}{lV{2}}{}     & \multicolumn{1}{lV{2}}{\rotatebox{90}{pass@1~}}   & \multicolumn{1}{lV{2}}{\rotatebox{90}{pass@2~}}     & \multicolumn{1}{lV{2}}{\rotatebox{90}{pass@3~}}     & \multicolumn{1}{lV{2}}{\rotatebox{90}{pass@1~}}    & \multicolumn{1}{lV{2}}{\rotatebox{90}{pass@2~}}  & \multicolumn{1}{lV{2}}{\rotatebox{90}{pass@3~}}    \bigstrut\\ \clineB{3-8}{2}
&   & \multicolumn{1}{l}{}     & \multicolumn{1}{l}{}     & \multicolumn{1}{l}{}     & \multicolumn{1}{l}{}     & \multicolumn{1}{l}{}     & \multicolumn{1}{l}{}         \bigstrut\\[-1.2em] \hlineB{2}
\multicolumn{1}{V{2}lV{2}}{} &
 \multicolumn{1}{lV{2}}{\textsc{\textsc{Baseline}}} &
  \multicolumn{1}{rV{2}}{183} &
  \multicolumn{1}{rV{2}}{207} &
  \multicolumn{1}{rV{2}}{218} &
  \multicolumn{1}{rV{2}}{$\cdot$} &
  \multicolumn{1}{rV{2}}{$\cdot$} &
  \multicolumn{1}{rV{2}}{$\cdot$} \bigstrut\\ \cline{2-2}
\multicolumn{1}{V{2}lV{2}}{\multirow{-2}{*}{Gpt4o}} &
  \multicolumn{1}{lV{2}}{\sysname{}} &
  \multicolumn{1}{rV{2}}{\cellcolor{orange!10}204} &
  \multicolumn{1}{rV{2}}{\cellcolor{orange!10}228} &
  \multicolumn{1}{rV{2}}{\cellcolor{orange!10}233} &
  \multicolumn{1}{rV{2}}{\cellcolor{blue!10}11\%} &
  \multicolumn{1}{rV{2}}{\cellcolor{blue!10}10\%} &
  \multicolumn{1}{rV{2}}{\cellcolor{blue!10}7\%}\bigstrut\\ \hline
\multicolumn{1}{V{2}lV{2}}{} &
 \multicolumn{1}{lV{2}}{\textsc{\textsc{Baseline}}} &
  \multicolumn{1}{rV{2}}{143} &
  \multicolumn{1}{rV{2}}{158} &
  \multicolumn{1}{rV{2}}{167} &
  \multicolumn{1}{rV{2}}{$\cdot$} &
  \multicolumn{1}{rV{2}}{$\cdot$} &
  \multicolumn{1}{rV{2}}{$\cdot$} \bigstrut\\ \cline{2-2}
\multicolumn{1}{V{2}lV{2}}{\multirow{-2}{*}{Gpt4o-mini}} &
  \multicolumn{1}{lV{2}}{\sysname{}} &
  \multicolumn{1}{rV{2}}{\cellcolor{orange!10}158} &
  \multicolumn{1}{rV{2}}{\cellcolor{orange!10}177} &
  \multicolumn{1}{rV{2}}{\cellcolor{orange!10}179} &
  \multicolumn{1}{rV{2}}{\cellcolor{blue!10}10\%} &
  \multicolumn{1}{rV{2}}{\cellcolor{blue!10}12\%} &
  \multicolumn{1}{rV{2}}{\cellcolor{blue!10}7\%}\bigstrut\\ \hline
\multicolumn{1}{V{2}lV{2}}{} &
 \multicolumn{1}{lV{2}}{\textsc{\textsc{Baseline}}} &
  \multicolumn{1}{rV{2}}{238} &
  \multicolumn{1}{rV{2}}{246} &
  \multicolumn{1}{rV{2}}{250} &
  \multicolumn{1}{rV{2}}{$\cdot$} &
  \multicolumn{1}{rV{2}}{$\cdot$} &
  \multicolumn{1}{rV{2}}{$\cdot$} \bigstrut\\ \cline{2-2}
\multicolumn{1}{V{2}lV{2}}{\multirow{-2}{*}{Claude}} &
  \multicolumn{1}{lV{2}}{\sysname{}} &
  \multicolumn{1}{rV{2}}{\cellcolor{orange!10}243} &
  \multicolumn{1}{rV{2}}{\cellcolor{orange!10}262} &
  \multicolumn{1}{rV{2}}{\cellcolor{orange!10}267} &
  \multicolumn{1}{rV{2}}{\cellcolor{blue!10}2\%} &
  \multicolumn{1}{rV{2}}{\cellcolor{blue!10}6\%} &
  \multicolumn{1}{rV{2}}{\cellcolor{blue!10}7\%}\bigstrut\\ \hline
\multicolumn{1}{V{2}lV{2}}{} &
 \multicolumn{1}{lV{2}}{\textsc{\textsc{Baseline}}} &
  \multicolumn{1}{rV{2}}{161} &
  \multicolumn{1}{rV{2}}{177} &
  \multicolumn{1}{rV{2}}{184} &
  \multicolumn{1}{rV{2}}{$\cdot$} &
  \multicolumn{1}{rV{2}}{$\cdot$} &
  \multicolumn{1}{rV{2}}{$\cdot$} \bigstrut\\ \cline{2-2}
\multicolumn{1}{V{2}lV{2}}{\multirow{-2}{*}{Gemini}} &
  \multicolumn{1}{lV{2}}{\sysname{}} &
  \multicolumn{1}{rV{2}}{\cellcolor{orange!10}176} &
  \multicolumn{1}{rV{2}}{\cellcolor{orange!10}198} &
  \multicolumn{1}{rV{2}}{\cellcolor{orange!10}203} &
  \multicolumn{1}{rV{2}}{\cellcolor{blue!10}9\%} &
  \multicolumn{1}{rV{2}}{\cellcolor{blue!10}12\%} &
  \multicolumn{1}{rV{2}}{\cellcolor{blue!10}10\%}\bigstrut\\ \hline
\multicolumn{1}{V{2}lV{2}}{} &
 \multicolumn{1}{lV{2}}{\textsc{\textsc{Baseline}}} &
  \multicolumn{1}{rV{2}}{121} &
  \multicolumn{1}{rV{2}}{149} &
  \multicolumn{1}{rV{2}}{165} &
  \multicolumn{1}{rV{2}}{$\cdot$} &
  \multicolumn{1}{rV{2}}{$\cdot$} &
  \multicolumn{1}{rV{2}}{$\cdot$} \bigstrut\\ \cline{2-2}
\multicolumn{1}{V{2}lV{2}}{\multirow{-2}{*}{Llama}} &
  \multicolumn{1}{lV{2}}{\sysname{}} &
  \multicolumn{1}{rV{2}}{\cellcolor{orange!10}134} &
  \multicolumn{1}{rV{2}}{\cellcolor{orange!10}168} &
  \multicolumn{1}{rV{2}}{\cellcolor{orange!10}176} &
  \multicolumn{1}{rV{2}}{\cellcolor{blue!10}11\%} &
  \multicolumn{1}{rV{2}}{\cellcolor{blue!10}13\%} &
  \multicolumn{1}{rV{2}}{\cellcolor{blue!10}7\%}\bigstrut\\ \hline
\multicolumn{1}{V{2}lV{2}}{} &
 \multicolumn{1}{lV{2}}{\textsc{\textsc{Baseline}}} &
  \multicolumn{1}{rV{2}}{126} &
  \multicolumn{1}{rV{2}}{152} &
  \multicolumn{1}{rV{2}}{166} &
  \multicolumn{1}{rV{2}}{$\cdot$} &
  \multicolumn{1}{rV{2}}{$\cdot$} &
  \multicolumn{1}{rV{2}}{$\cdot$} \bigstrut\\ \cline{2-2}
\multicolumn{1}{V{2}lV{2}}{\multirow{-2}{*}{Mistral}} &
  \multicolumn{1}{lV{2}}{\sysname{}} &
  \multicolumn{1}{rV{2}}{\cellcolor{orange!10}159} &
  \multicolumn{1}{rV{2}}{\cellcolor{orange!10}176} &
  \multicolumn{1}{rV{2}}{\cellcolor{orange!10}179} &
  \multicolumn{1}{rV{2}}{\cellcolor{blue!10}26\%} &
  \multicolumn{1}{rV{2}}{\cellcolor{blue!10}16\%} &
  \multicolumn{1}{rV{2}}{\cellcolor{blue!10}8\%}\bigstrut\\ \hlineB{2}
\end{tabular}}

\end{table*}

\subsection*{RQ1.~Overall Performance}
In this section, we evaluate \sysname{} against a baseline that does not use specifications.
% This allows us to evaluate our multi-modal specifications that work cohesively to produce better translations.

\noindent
\textbf{Evaluation:} We measure the number of accurate translations produced by \sysname{} (at various steps) against the baseline (where the prompt contains no additional specs) at various \textit{pass@k}. Note that each pass at k corresponds to the equivalent step-k, \ie pass@1 corresponds to step 1, and so on.
% We draw this distinction in nomenclature so as to highlight the the impact of each specification as opposed to resampling the language model with no extra specifications provided to it.
We also measure \% improvement over the baseline at various steps/pass@k's. In this comparison, a positive percentage score indicates that, compared to the baseline, more source programs are accurately translated by \sysname{}. Our findings are summarized in \Cref{tab:c2rust-and-c2go,tab:js2ts}. 

\noindent
\textbf{Discussion:} The tabulated findings may be summarized as below:
\bi
\item \textit{\sysname{} produces more correct translations compared to baselines}: \sysname{} generally shows improvement over the baseline. The largest relative gains are on the C to Rust translation task, and in particular, Mistral shows relative improvements of between \textbf{31\%} and \textbf{46\%}. The best-performing base models are GPT4o and Claude, but they too are able to benefit from using \sysname{}.

% While GPT4o and Claude produce larger number of correct translations in both C to Go and C to Rust, Granite 34b outperforms equivalently sized Deepseek and the much larger GPT3.5 in terms of the number of correct translations for C to Rust. And Deepseek outperforms GPT3.5 and Granite in translating C to Go. \todo{I have commented out your previous paragraphs on absolute number, please list a few noteworthy ones here like +23\% in step-2 gemini etc.}

% \todo{I have commented out your previous paragraphs on absolute number, please list a few noteworthy ones here like +26\% in step-2 gemini etc.}

\item \textit{The largest relative increases generally occur for models that have the lowest baseline performances, and vice-versa}: For example, Mistral and Llama are the two worst models at the C to Rust task, and the highest ``improvement \%'' on this task is on Mistral (31-46\%) followed by Llama (23\%). The converse is also true - Claude is the best performing model on C to Go, and its performance gain (2-7\%) is the lowest of all models on this task.

\begin{table}[t!]
\caption{Comparing \sysname{} to the baseline for Javascript to Typescript. The best absolute numbers for each column are in \colorbox{orange!10}{orange}, percentage increases are in \colorbox{blue!10}{blue}, percentage decreases are in \colorbox{red!10}{pink}. \sysname{} does improve performance, but gains are limited.
% \todo{Please add a one
% line summary of the colors and findings. Also note that DEEPSEEK is missing (you'll find the lastest results you'd asked for in the repo), please uncomment the lines corresponding these in that last two rows and update them. make sure to use \textbackslash hlineB\{2\} to get that thick bottom rule line.}
}
\label{tab:js2ts}
\centering
\resizebox{0.95\linewidth}{!}{
\begin{tabular}{llrrrrrr}
\clineB{3-8}{2}
 & \multicolumn{1}{lV{2}}{}    & \multicolumn{3}{c|}{No. correct}  & \multicolumn{3}{c|}{improvement \%}  \bigstrut\\ \clineB{3-8}{2}
& \multicolumn{1}{lV{2}}{}     & \multicolumn{1}{lV{2}}{\rotatebox{90}{pass@1~}}   & \multicolumn{1}{lV{2}}{\rotatebox{90}{pass@2~}}     & \multicolumn{1}{lV{2}}{\rotatebox{90}{pass@3~}}     & \multicolumn{1}{lV{2}}{\rotatebox{90}{pass@1~}}    & \multicolumn{1}{lV{2}}{\rotatebox{90}{pass@2~}}  & \multicolumn{1}{lV{2}}{\rotatebox{90}{pass@3~}}    \bigstrut\\ \clineB{3-8}{2}
&   & \multicolumn{1}{l}{}     & \multicolumn{1}{l}{}     & \multicolumn{1}{l}{}     & \multicolumn{1}{l}{}     & \multicolumn{1}{l}{}     & \multicolumn{1}{l}{}         \bigstrut\\[-1.2em] \hlineB{2}
\multicolumn{1}{V{2}lV{2}}{} &
 \multicolumn{1}{lV{2}}{\textsc{\textsc{Baseline}}} &
  \multicolumn{1}{rV{2}}{\cellcolor{orange!10}289} &
  \multicolumn{1}{rV{2}}{\cellcolor{orange!10}295} &
  \multicolumn{1}{rV{2}}{\cellcolor{orange!10}295} &
  \multicolumn{1}{rV{2}}{$\cdot$} &
  \multicolumn{1}{rV{2}}{$\cdot$} &
  \multicolumn{1}{rV{2}}{$\cdot$} \bigstrut\\ \cline{2-2}
\multicolumn{1}{V{2}lV{2}}{\multirow{-2}{*}{Gpt4o}} &
  \multicolumn{1}{lV{2}}{\sysname{}} &
  \multicolumn{1}{rV{2}}{284} &
  \multicolumn{1}{rV{2}}{292} &
  \multicolumn{1}{rV{2}}{294} &
  \multicolumn{1}{rV{2}}{\cellcolor{red!10}-2\%} &
  \multicolumn{1}{rV{2}}{\cellcolor{red!10}-1\%} &
  \multicolumn{1}{rV{2}}{\cellcolor{red!10}-0.3\%}\bigstrut\\ \hline
\multicolumn{1}{V{2}lV{2}}{} &
 \multicolumn{1}{lV{2}}{\textsc{\textsc{Baseline}}} &
  \multicolumn{1}{rV{2}}{\cellcolor{orange!10}282} &
  \multicolumn{1}{rV{2}}{284} &
  \multicolumn{1}{rV{2}}{285} &
  \multicolumn{1}{rV{2}}{$\cdot$} &
  \multicolumn{1}{rV{2}}{$\cdot$} &
  \multicolumn{1}{rV{2}}{$\cdot$} \bigstrut\\ \cline{2-2}
\multicolumn{1}{V{2}lV{2}}{\multirow{-2}{*}{Gpt4o-mini}} &
  \multicolumn{1}{lV{2}}{\sysname{}} &
  \multicolumn{1}{rV{2}}{279} &
  \multicolumn{1}{rV{2}}{\cellcolor{orange!10}289} &
  \multicolumn{1}{rV{2}}{\cellcolor{orange!10}289} &
  \multicolumn{1}{rV{2}}{\cellcolor{red!10}-1\%} &
  \multicolumn{1}{rV{2}}{\cellcolor{blue!10}2\%} &
  \multicolumn{1}{rV{2}}{\cellcolor{blue!10}1\%}\bigstrut\\ \hline
\multicolumn{1}{V{2}lV{2}}{} &
 \multicolumn{1}{lV{2}}{\textsc{\textsc{Baseline}}} &
  \multicolumn{1}{rV{2}}{282} &
  \multicolumn{1}{rV{2}}{285} &
  \multicolumn{1}{rV{2}}{286} &
  \multicolumn{1}{rV{2}}{$\cdot$} &
  \multicolumn{1}{rV{2}}{$\cdot$} &
  \multicolumn{1}{rV{2}}{$\cdot$} \bigstrut\\ \cline{2-2}
\multicolumn{1}{V{2}lV{2}}{\multirow{-2}{*}{Claude}} &
  \multicolumn{1}{lV{2}}{\sysname{}} &
  \multicolumn{1}{rV{2}}{\cellcolor{orange!10}283} &
  \multicolumn{1}{rV{2}}{\cellcolor{orange!10}292} &
  \multicolumn{1}{rV{2}}{\cellcolor{orange!10}292} &
  \multicolumn{1}{rV{2}}{\cellcolor{blue!10}0.4\%} &
  \multicolumn{1}{rV{2}}{\cellcolor{blue!10}2\%} &
  \multicolumn{1}{rV{2}}{\cellcolor{blue!10}2\%}\bigstrut\\ \hline
\multicolumn{1}{V{2}lV{2}}{} &
 \multicolumn{1}{lV{2}}{\textsc{\textsc{Baseline}}} &
  \multicolumn{1}{rV{2}}{\cellcolor{orange!10}284} &
  \multicolumn{1}{rV{2}}{289} &
  \multicolumn{1}{rV{2}}{290} &
  \multicolumn{1}{rV{2}}{$\cdot$} &
  \multicolumn{1}{rV{2}}{$\cdot$} &
  \multicolumn{1}{rV{2}}{$\cdot$} \bigstrut\\ \cline{2-2}
\multicolumn{1}{V{2}lV{2}}{\multirow{-2}{*}{Gemini}} &
  \multicolumn{1}{lV{2}}{\sysname{}} &
  \multicolumn{1}{rV{2}}{282} &
  \multicolumn{1}{rV{2}}{\cellcolor{orange!10}291} &
  \multicolumn{1}{rV{2}}{\cellcolor{orange!10}292} &
  \multicolumn{1}{rV{2}}{\cellcolor{red!10}-1\%} &
  \multicolumn{1}{rV{2}}{\cellcolor{blue!10}1\%} &
  \multicolumn{1}{rV{2}}{\cellcolor{blue!10}1\%}\bigstrut\\ \hline
\multicolumn{1}{V{2}lV{2}}{} &
 \multicolumn{1}{lV{2}}{\textsc{\textsc{Baseline}}} &
  \multicolumn{1}{rV{2}}{268} &
  \multicolumn{1}{rV{2}}{278} &
  \multicolumn{1}{rV{2}}{283} &
  \multicolumn{1}{rV{2}}{$\cdot$} &
  \multicolumn{1}{rV{2}}{$\cdot$} &
  \multicolumn{1}{rV{2}}{$\cdot$} \bigstrut\\ \cline{2-2}
\multicolumn{1}{V{2}lV{2}}{\multirow{-2}{*}{Llama}} &
  \multicolumn{1}{lV{2}}{\sysname{}} &
  \multicolumn{1}{rV{2}}{\cellcolor{orange!10}270} &
  \multicolumn{1}{rV{2}}{\cellcolor{orange!10}284} &
  \multicolumn{1}{rV{2}}{\cellcolor{orange!10}285} &
  \multicolumn{1}{rV{2}}{\cellcolor{blue!10}1\%} &
  \multicolumn{1}{rV{2}}{\cellcolor{blue!10}2\%} &
  \multicolumn{1}{rV{2}}{\cellcolor{blue!10}1\%}\bigstrut\\ \hline
\multicolumn{1}{V{2}lV{2}}{} &
 \multicolumn{1}{lV{2}}{\textsc{\textsc{Baseline}}} &
  \multicolumn{1}{rV{2}}{\cellcolor{orange!10}285} &
  \multicolumn{1}{rV{2}}{289} &
  \multicolumn{1}{rV{2}}{289} &
  \multicolumn{1}{rV{2}}{$\cdot$} &
  \multicolumn{1}{rV{2}}{$\cdot$} &
  \multicolumn{1}{rV{2}}{$\cdot$} \bigstrut\\ \cline{2-2}
\multicolumn{1}{V{2}lV{2}}{\multirow{-2}{*}{Mistral}} &
  \multicolumn{1}{lV{2}}{\sysname{}} &
  \multicolumn{1}{rV{2}}{280} &
  \multicolumn{1}{rV{2}}{\cellcolor{orange!10}291} &
  \multicolumn{1}{rV{2}}{\cellcolor{orange!10}292} &
  \multicolumn{1}{rV{2}}{\cellcolor{red!10}-2\%} &
  \multicolumn{1}{rV{2}}{\cellcolor{blue!10}1\%} &
  \multicolumn{1}{rV{2}}{\cellcolor{blue!10}1\%}\bigstrut\\ \hlineB{2}
\end{tabular}}
\end{table}

\item \textit{\sysname{} also helps improve Javascript to Typescript translation albeit with considerably less pronounced gains compared to the other two translation tasks}: \sysname{} generally improves results over the baseline (\Cref{tab:js2ts}), with consistent percentage improvements for Claude and Llama. However, overall improvements are far less pronounced compared to other translation tasks. We conjecture that this is due to the high degree of semantic and syntactic similarity between the two languages. JavaScript and TypeScript share many features and structures, and the translation process is less reliant on detailed specifications to achieve correctness.
% This similarity reduces the need for extensive specification guidance, as
The LLM can infer the correct translations based on existing structural and semantic parallels. We also see in \Cref{tab:js2ts} that specifications can be detrimental in certain cases. On manual inspection, we find that the JavaScript specifications are extremely verbose \footnote{Since JavaScript does not require procedural code to be enclosed within functions, we ask the LLM to partition the code into logically coherent segments. In practice, this leads to as many as 20 segments in some programs, each with their own preconditions and postconditions.}, and add a lot of tokens to the input. As the length of the context increases, LLMs are more prone to hallucinations and other errors \cite{peng2023yarn}. Future work could explore generating concise specifications for JavaScript.
\ei

\begin{figure}
    \centering
    \includegraphics[width=0.8\columnwidth]{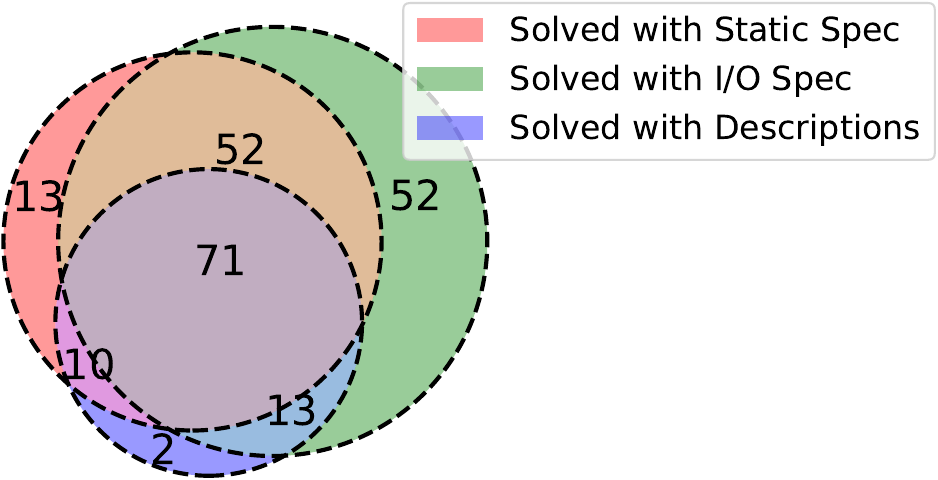}
    \caption{Our different specification modalities are complementary. This diagram shows how many out of the 300 C to Rust translations are solved using each individual specification modality. All results are with \texttt{gpt-4o}. Diagrams of other language pairs are in the Supplementary Material.}
    \label{fig:venn}
\end{figure}

\subsection*{RQ2.~Impact of Individual Specification Modalities} 
Here we evaluate the efficacy of each specification modality individually. Our hypothesis is that different modalities would act in a complementary fashion, whereby each modality might provide some extra information for translation that the other modalities do not provide.

\noindent
\textbf{Evaluation:} For each of the 300 C and JavaScript programs, we attempt to perform C to Rust, C to Go, and JavaScript to TypeScript translation using \texttt{gpt-4o}. While generating translations, we provide each specification modality individually along with the program and obtain one candidate translation per modality. 

\noindent
\textbf{Discussion:} We evaluate all the translations and plot a Venn Diagram in \Cref{fig:venn} to visualize the contributions of each individual specification modality. Because of a lack of space, we show only the C to Rust diagram here; the others are in the appendix. There are 71 programs that are solved with all three modalities, showing that there is a substantial overlap. However, the remaining 213 programs are solved with some combination of two or fewer modalities, showing that the different modalities are indeed complementary.~\res{elaborate the results with numbers.}

% \todo{I don't understand what the next few lines are saying, could you please rephrase? Also, we shouldn avoid new math symbols in the results section, could we just use text unless it is for an abbreviation.}

% We compare this with the performance of $\mathcal{P}_\text{trans}$ for those same programs, again using \texttt{gpt-4o}. We repeat the same process for $\mathcal{P}^\mathcal{T}_\text{trans}$ and $\mathcal{P}^\mathcal{D}_\text{trans}$.

% To evaluate the effect of self-consistency, we also pick a random candidate specification that has \textit{not} been validated against the corresponding program and therefore has no guarantee of being correct. We try to use this specification for translation.

% \todo{1. which table is this? Did I accidentally remove it? Please reinsert if so. 2. We can be more specific here on what we mean by improved quality with numbers.}
% The results are shown in \Cref{tab:individual}. We can see that each mode of specification improves translation quality against the \textsc{\textsc{\textsc{Baseline}}}, for both Rust and Go.\todo{please comment on javascript}

\subsection*{RQ3.~Impact of Design Choices}
\label{sect:design-choices}
In this RQ, we first investigate the effectiveness of our multi-stage approach in \sysname{}, where we provide each of the modalities of specifications one by one, as opposed to providing all of them together. Next, we evaluate the impact of the self-consistency check on translation quality.

\noindent
\textbf{Evaluation 1:}  We take a single setting - GPT-4o for C to Rust translation, and compare \sysname{} with a baseline where we provide all the specifications together in the same prompt, and take the top $k$ candidates to compute pass@k. The results are shown in \Cref{tab:combined_ablation}.

\noindent
\textbf{Discussion 1:}  We can see that \sysname{} is better than the combined baseline. This justifies our intuition in \Cref{sect:stage-3} that using all modalities simultaneously can create a surfeit of information, expanding the size of the LLM input and degrading its performance. ~\res{Why? Give some insights}

\noindent
\textbf{Evaluation 2:} We again take GPT-4o for C to Rust translation, and compare \sysname{} with a baseline where we take the \textit{first} specification generated by our specification generation model, with no correctness check. We call this a ``one-shot specification''. The results are shown in \Cref{tab:sc_ablation}.

\noindent
\textbf{Discussion 2:} The self-consistency check makes a noticeable positive difference for translation using Static and I/O specifications. In fact, for I/O specifications, translation with one-shot specifications is \textit{worse} than translation with \textit{no} specifications. However, translation with Descriptions is unaffected by the self-consistency check. We hypothesize that this is because of the imprecise nature of natural language. The inherent ambiguity of the specification could mask small semantic mistakes. Likewise, this same ambiguity could be a threat to the validity of the self-consistency filter for descriptions.

\begin{table}[t]
    \caption{Evaluating design choices. All results for GPT-4o on C to Rust. The best performing method in each column is highlighted in \colorbox{orange!10}{orange} for absolute numbers and \colorbox{blue!10}{blue} for percentages.}
    \begin{subtable}[t]{\columnwidth}
    \resizebox{0.95\linewidth}{!}{
    \begin{tabular}{l|c|c|c|c|c|c|}
    \clineB{2-7}{2}
    & \rotatebox{90}{pass@1~} & \rotatebox{90}{pass@2~} & \rotatebox{90}{pass@3~} & \rotatebox{90}{pass@1~} & \rotatebox{90}{pass@2~} & \rotatebox{90}{pass@3~} 
    \bigstrut\\ \clineB{2-7}{2}
\multicolumn{1}{l}{} &   \multicolumn{1}{l}{}     & \multicolumn{1}{l}{}     & \multicolumn{1}{l}{}     & \multicolumn{1}{l}{}     & \multicolumn{1}{l}{}     & \multicolumn{1}{l}{}         \bigstrut\\[-1.2em] \hlineB{2}
    \multicolumn{1}{|l|}{w/o Specifications} & 176 & 192 & 202 & $\cdot$ & $\cdot$ & $\cdot$ \\[0.25em]
    \clineB{1-1}{1}
    \multicolumn{1}{|l|}{All Specs Together} & 195 & 215 & 219 & 11\% & 12\% & 8\% \\[0.25em]
    \clineB{1-1}{1}
    \multicolumn{1}{|l|}{\textsc{\sysname{}}} & \cellcolor{orange!10} 196 & \cellcolor{orange!10} 219 & \cellcolor{orange!10} 220 & \cellcolor{blue!10} 11\% & \cellcolor{blue!10} 14\% & \cellcolor{blue!10} 9\%
    \bigstrut \\[0.25em] \hlineB{2}
    \end{tabular}}
    \vspace{3pt}
    \caption{Assessing the impact of providing all specification modalities together, instead of in multiple stages, \ie \sysname{}.  Combining modalities is better than no specifications, but \sysname{} is better than both of them.}
    \label{tab:combined_ablation}
    \end{subtable}

    \begin{subtable}[t]{\columnwidth}
    \resizebox{0.95\linewidth}{!}{
    \begin{tabular}{l|c|c|c|c|c|c|}
    \clineB{2-7}{2}
    & \rotatebox{90}{\textsc{Static}~} & \rotatebox{90}{\textsc{I/O}~} & \rotatebox{90}{\textsc{Desc}~} & \rotatebox{90}{\textsc{Static}~} & \rotatebox{90}{\textsc{I/O}~} & \rotatebox{90}{\textsc{Desc}~} 
    \bigstrut\\ \clineB{2-7}{2}
\multicolumn{1}{l}{} &   \multicolumn{1}{l}{}     & \multicolumn{1}{l}{}     & \multicolumn{1}{l}{}     & \multicolumn{1}{l}{}     & \multicolumn{1}{l}{}     & \multicolumn{1}{l}{}         \bigstrut\\[-1.2em] \hlineB{2}
    \multicolumn{1}{|l|}{\# Programs w. Spec} & 189 & 294 & 134 & $\cdot$ & $\cdot$ & $\cdot$ \bigstrut\\
    \hlineB{2}
\multicolumn{1}{l}{} &   \multicolumn{1}{l}{}     & \multicolumn{1}{l}{}     & \multicolumn{1}{l}{}     & \multicolumn{1}{l}{}     & \multicolumn{1}{l}{}     & \multicolumn{1}{l}{}         \bigstrut\\[-1.2em] \hlineB{2}
    
    \multicolumn{1}{|l|}{w/o Specifications} & 131 & 174 & 88 & $\cdot$ & $\cdot$ & $\cdot$ \\[0.25em]
    \clineB{1-1}{1}
    \multicolumn{1}{|l|}{One-shot Spec} & 140 & 170 & \cellcolor{orange!10} 97 & 7\% & -2\% & \cellcolor{blue!10} 10\% \\[0.25em]
    \clineB{1-1}{1}
    \multicolumn{1}{|l|}{Self-consistent Spec} & \cellcolor{orange!10} 146 & \cellcolor{orange!10} 188 & 96 & \cellcolor{blue!10} 11\% & \cellcolor{blue!10} 8\% &  9\%
    \bigstrut \\[0.25em] \hlineB{2}
    \end{tabular}}
    \vspace{3pt}
    \caption{Assessing the impact of the self-consistency check. ``One-shot Spec'' refers to translation using the first generated specification, without checking for self-consistency. For Static and I/O Specs, the self-consistency check makes a noticeable positive difference.}
    \label{tab:sc_ablation}
    \end{subtable}
\end{table}

\subsection*{RQ4.~Evaluating the Generated Specifications}

We would like to determine whether our specification generation process is able to generate high-quality self-consistent specifications. 

\noindent
\textbf{Evaluation:} To do this, we check how many programs in our benchmark dataset we are able to generate self-consistent specifications for. For each program, we generate up to $k=10$ candidate tests, and up to $k = 6$ static specifications and descriptions. In all the cases, we stop as soon as we find a self-consistent specification. We use \texttt{gpt-4o} as our LLM for both the ``forward'' specification generation as well as the ``reverse'' code generation process. We also measured the generated test coverage for C programs using the \texttt{gcov} tool.

\begin{table}[]
    \centering
    \small
    \caption{The number of source programs for which we are able to generate a successful specification.}
    \label{tab:specgen_success}
    \begin{tabular}{c|c|c|c|c}
    \cline{1-5}
        \textbf{Lang} & \textbf{Total} & \textbf{Static} & \textbf{I/O} & \textbf{Desc} \bigstrut\\
        \cline{1-5}
        C & 300 & \makecell{189\bigstrut~({63.0\%})} & \makecell{294\bigstrut~({98.0\%})} & \makecell{134\bigstrut~ ({44.7\%})}\bigstrut\\
         \cline{1-5}
        JS & 300 & \makecell{199\bigstrut~({66.3\%})} & \makecell{296\bigstrut~({98.7\%})} & \makecell{189\bigstrut~({63.0\%})} \bigstrut\\
    \cline{1-5}
    \end{tabular}
\end{table}
\noindent
\textbf{Discussion:} The number of self-consistent specifications generated by \texttt{gpt-4o} of each category is shown in \Cref{tab:specgen_success}.
% We find that \texttt{gpt-4o} is able to generate self-consistent static specifications for \textbf{$\sim$60\%}, self-consistent tests for \textbf{$\sim$98\%}, and self-consistent descriptions for \textbf{$\sim$40\%} of the programs.
% A box plot of the number of attempts needed to generate a successful specification is shown in \Cref{fig:boxplot}
The average coverage was \textbf{91.5\%}, which is an indicator that the tests are of high quality. For the self-consistent descriptions and static specifications, the fact that we are able to recover a functioning program from them is a strong validation of their quality. We also manually inspected several of them and confirmed that they are a good representation of the program's behavior.

\section{Case Study of Translating Projects}
\label{sec:results_fp}

\begin{table*}
\centering
\resizebox{0.85\textwidth}{!}{
\begin{tabular}{c}
\toprule
\begin{subtable}[b]{0.95\columnwidth}
\lstset{style=myc, basicstyle=\small\ttfamily, linewidth=\columnwidth}
\begin{lstlisting}
/* <@\textbf{\textcolor{greencomments}{Static Spec:}}@>
Precondition:
1. `fd` must be a valid file descriptor.
...
Postcondition:
1. The function attempts to read/write up to `count` bytes from/to the file descriptor `fd` using `safe_rw`.
2. The return value is the total number of bytes successfully read/written (`total`).
... */
/* <@\textbf{\textcolor{greencomments}{I/O Spec:}}@>
Arguments: ["1","processor\t: 0<@\textcolor{greencomments}{\textdollar}@>\nvendor_id...","77046"]
Return: "77046" */
# define full_rw full_write
size_t full_rw (int fd, const void *buf, size_t count){
  size_t total = 0;
  const char *ptr = (const char *) buf;
  while (count > 0){
      size_t n_rw = safe_rw (fd, ptr, count);
      if (n_rw == (size_t) - 1)
        break;
      if (n_rw == 0){
          errno = ZERO_BYTE_TRANSFER_ERRNO;
          break;
      }
      total += n_rw;
      ptr += n_rw;
      count -= n_rw;
  }
  return total;
}
\end{lstlisting}
\caption{A C function \TT{full\_write} from the \TT{cat} utility, along with its generated static and I/O specifications.}
\end{subtable}
\hspace{1cm}
\begin{subtable}[b]{0.95\columnwidth}
\lstset{language=Rust, frame=none, style=colouredRust, basicstyle=\small\ttfamily, escapeinside={<@}{@>}, breaklines=true}
\begin{lstlisting}
fn full_write_rust(fd: RawFd, buf: &[u8]) -> usize {
  let mut total_written = 0;
  let mut remaining = buf;
  while !remaining.is_empty() {
    match safe_rw(fd, remaining) {
      Ok(bytes_written) => {
        if bytes_written == 0 {
          break;
        }
        total_written += bytes_written;
        remaining = &remaining[bytes_written..];
      }
      Err(_) => {
        // Break on error (safe_rw handles setting errno)
        break;
      }
    }
  }
  total_written
}

fn safe_rw(fd: RawFd, buf: &[u8]) -> io::Result<usize> {
  use std::{fs::File, os::unix::io::FromRawFd};
  
  // Wrap the raw file descriptor in a File for safe I/O
  let mut file = unsafe { File::from_raw_fd(fd) };
  let result = file.write(buf);
  // Prevent the file from being closed when it goes out of scope
  std::mem::forget(file);
  result
}
\end{lstlisting}
\caption{The translated Rust version generated by \sysname{}, consisting of the safe function \TT{full\_write\_rust} and a helper function \TT{safe\_rw}.}
\end{subtable}\\
\bottomrule
\end{tabular}%
}
\vspace{0.3cm}
\caption{An example of translating a single function from a large C project into Rust using \sysname{}. The LLM, GPT-4o, is able to use the information from these specifications to augment its understanding of the function, thereby producing an accurate translation.}%
\label{tab:full_proj_eg}
\end{table*}

\subsection{Approach}
\label{sect:full-project}

We would like to adapt our specification-driven approach to translate full software projects. %, consisting of multiple files with several hundreds of lines of code. 
This setting presents multiple challenges:

\noindent
\circled{1} \textit{The limitation of the LLM context window}. The entire project may consist of thousands of lines of code, so it might not fit into an LLM input context window.

\noindent
\textbf{Solution:} We need to decompose the project into smaller pieces that can be individually translated. In the competitive coding setting, our Static and NL specifications were generated at the function level, so here too, we generate specifications at the granularity of each function.

\noindent
\circled{2} \textit{How to evaluate the correctness of each individual function translation}. If a function is translated from one language to another, the code may no longer compile because of other functions that call this function. Further, although we may have \textit{end-to-end} test cases, it is difficult to evaluate a \textit{standalone} function translation.

\noindent
\textbf{Solution:} Our solution is to use cross-language foreign function APIs to enable functions in one language to call functions in another language, and vice-versa. That way, the entire code still compiles and runs end-to-end. The translated function becomes a modified link in a chain of functions, and if we run end-to-end tests, we can see whether the modified function preserves the overall program's correctness with respect to these tests.

\noindent
\circled{3} \textit{The kinds of errors that LLMs make while translating complex functions with many dependencies}. There are inevitably some errors due to missing imports, redeclaring variables, and other ``shallow'' compilation errors that are \textit{not} due to lack of understanding of function semantics.

\noindent
\textbf{Solution:} We perform a finite number of iterations of compiler feedback-aided repair, where we take the compiler log and give it to the LLM to re-generate its solution. We do this both for the baseline as well as for \sysname{}. 

\noindent
\circled{4} \textit{How to curate different kinds of specifications in this setting}. Generating specifications for individual functions using an LLM is easy, but \textit{validating} these using self-consistency requires a different approach to that presented in \Cref{sect:stage-2}.

\noindent
\textbf{Solution:} To generate \textit{static specifications} for a single function, we ask the LLM to generate a precondition and postcondition. Then, using this precondition and postcondition, we generate a back-translation as in \Cref{sect:stage-2}, but only of this single function. We \textit{swap out} the original version and \textit{swap in} the back-translated version, and run end-to-end tests to verify the correctness of the back-translation.

\noindent
We could follow a similar process as above for \textit{NL specifications}, but in the project we chose, most functions were already annotated with a docstring that briefly described their behavior. We extracted these docstrings and used them as our NL specifications.

\noindent
When it comes to \textit{I/O specifications}, program-level or end-to-end test cases are unlikely to be of much assistance to the LLM in a full project setting. If we are translating a single function in the middle of a chain of function calls, then it is extremely unlikely that end-to-end tests can provide information to aid in translating this single function. Our solution is to instrument the entire program to log the arguments and return values of each function. We run the provided end-to-end tests and collect these intermediate values, which we then provide to the LLM as I/O specifications. For this paper, we use the provided end-to-end tests, but future work could investigate using fuzzing to generate tests.

% % of program $\mathcal{S}$, we generate an \textit{ordered} list of candidate translations $\left[\mathcal{S}^{stat}_\text{trans}, \mathcal{S}^{I/O}_\text{trans}, \mathcal{S}^{desc}_\text{trans}\right]$. If we were unable to generate any of $\pi_{stat}, \pi_{I/O}$ and $\pi_{desc}$, then we omit the corresponding translation from the list of candidates. So we have between zero and three candidate translations for each program.

% \begin{align*}
% \mathcal{S}^\text{stat}_\text{trans} &= \mathbb{M}~(~\lambda_\text{stat-trans}~(~\mathcal{S}, \pi_{stat}~)~)\\
% \mathcal{S}^\mathcal{T}_\text{trans} &= \mathbb{M}~(~\lambda_\text{test-trans}~(~\mathcal{S}, \pi_{I/O}~)~)\\
% \mathcal{S}^\mathcal{D}_\text{trans} &= \mathbb{M}~(~\lambda_\text{desc-trans}~(~\mathcal{S}, \pi_{desc}~)~)
% \end{align*}

% We hypothesize that the best translations are obtained using the most precise and complete specifications, which is why we use the invariants first, followed by the test cases, followed by the descriptions. Finally, we try a vanilla translation with no specifications, which is, in theory, the least likely to succeed. This conceptually corresponds to a pipeline, where we first try translation with one kind of specification, and if it doesn't work, move on to another kind of specification, and so on.

\subsection{Study Results}
%We would like to investigate whether \sysname{} can still be effective in a full-project setting, where there are several functions with complex inter-dependencies.

\textbf{Evaluation:} We run \sysname{} with suitable modifications as described in \Cref{sect:full-project} on our dataset consisting of 24 functions from \TT{cat}. We first generate specifications for each individual function, and then use these specifications to translate each function. After each translation, we verify correctness by running end-to-end tests. We measure pass@k in the same manner as in \Cref{sect:eval}, except for each individual function instead of each program. We also allow up to 3 rounds of compiler-feedback based LLM debugging before evaluating each candidate translation.

\textbf{Discussion:} We are able to generate self-consistent static specifications for \textbf{10} out of 24 functions. \textbf{18} out of 24 functions have docstrings, and \textbf{all} the functions have input-output specifications collected through instrumentation. \Cref{tab:full-project} shows the results of using these specifications to translate the 24 functions. We can see that \sysname{} yields a small improvement in pass@k for all values of k. Although this is a small delta, we expect that this will translate into a larger gain in terms of absolute results if we evaluate our technique on a larger dataset. Additionally, as we observed for standalone programs, using multiple specification modalities together is not as effective as our multi-stage pipeline. In fact, this actually degrades performance slightly compared to a baseline without specifications.

\Cref{tab:full_proj_eg} shows an example of a function in C translated to Rust, along with its static and I/O specifications. We can see that the generated specifications are semantically meaningful, and provide information that \sysname{} is able to effectively use the information from these specifications to guide its translations.

\begin{table}[t]
    \caption{Full-project translation results, consisting of 24 functions from the \TT{cat} utility. All results for GPT-4o on C to Rust. \sysname{} yields a small positive benefit to translation accuracy. Additionally, using all specification modalities together causes performance to degrade slightly.}
    \label{tab:full-project}
    \resizebox{0.9\linewidth}{!}{
    \begin{tabular}{l|c|c|c|c|c|c|}
    \clineB{2-7}{2}
    & \rotatebox{90}{pass@1~} & \rotatebox{90}{pass@2~} & \rotatebox{90}{pass@3~} & \rotatebox{90}{pass@1~} & \rotatebox{90}{pass@2~} & \rotatebox{90}{pass@3~} 
    \bigstrut\\ \clineB{2-7}{2}
\multicolumn{1}{l}{} &   \multicolumn{1}{l}{}     & \multicolumn{1}{l}{}     & \multicolumn{1}{l}{}     & \multicolumn{1}{l}{}     & \multicolumn{1}{l}{}     & \multicolumn{1}{l}{}         \bigstrut\\[-1.2em] \hlineB{2}
    \multicolumn{1}{|l|}{w/o Specifications} & 10 & 11 & 12 & $\cdot$ & $\cdot$ & $\cdot$ \\[0.25em]
    \clineB{1-1}{1}
    \multicolumn{1}{|l|}{All Specs Together} & 8 & 11 & 11 & -20\% & 0\% & -8\% \\[0.25em]
    \clineB{1-1}{1}    
    \multicolumn{1}{|l|}{\sysname{}} & \cellcolor{orange!10} 11 & \cellcolor{orange!10} 12 & \cellcolor{orange!10} 13 & \cellcolor{blue!10} 10\% & \cellcolor{blue!10} 9\% & \cellcolor{blue!10} 8\% 
    \bigstrut\\ \clineB{1-7}{2}
\multicolumn{1}{l}{} &   \multicolumn{1}{l}{}     & \multicolumn{1}{l}{}     & \multicolumn{1}{l}{}     & \multicolumn{1}{l}{}     & \multicolumn{1}{l}{}     & \multicolumn{1}{l}{}         \bigstrut\\[-1.2em] \hlineB{2}
    \multicolumn{1}{|l|}{Total \#Func.} & \multicolumn{3}{c|}{24} & $\cdot$ & $\cdot$ & $\cdot$ \bigstrut\\
    \hlineB{2}
    \end{tabular}%
    }
\end{table}

\section{Related Work}

% Our approach addresses the problem of code translation with LLMs, but also draws on a long line of work on using LLMs for generating program specifications.

\textbf{Code Translation with LLMs.} \citet{roziere2020unsupervised} proposed Trans-Coder, an early attempt to tackle code translation using transformer models. They proposed Back Translation, a fully unsupervised approach. \cite{ahmad2022summarize} refined Back Translation with the addition of a code summarization task. In contrast to unsupervised learning, supervised learning requires aligned parallel data across pairs of programming languages, which is challenging to obtain at scale. \cite{roziere2021leveraging} address this challenge by generating synthetic aligned data using another language model and filtering out incorrect translations using generated unit tests. The last few years have seen the development of large foundation models for code and natural language, that are pretrained on massive amounts of data with an autoregressive objective function. \citet{pan2024lost} and \citet{yang2024exploring} perform thorough empirical evaluations of the translation ability of these models. Recently, there has been a lot of interest in translating full software repositories \cite{shiraishi2024context,ibrahimzada2024repository,shetty2024syzygy,zhang2024scalable,nitin2025c2saferrust}. Although all of these works differ in their approach and setting, the only one that utilizes any kind of specifications is \citet{shetty2024syzygy}. They use carefully crafted Rust-specific properties collected through dynamic analysis to guide the LLM. Our specifications are much broader in scope, however.

\noindent
\textbf{Generating Program Specifications with LLMs.} The three kinds of program specifications that we consider in this paper are test cases, natural language descriptions, and static specifications. There have been a few recent papers that use LLMs to generate test cases for competitive coding problems. EvalPlus \cite{liu2023your} uses ChatGPT to generate test cases, and mutates them in order to rigorously test LLM-generated code solutions. CodeT \cite{chen2022codet} uses LLM-generated test cases to filter out code solutions from a list of LLM-generated candidate solutions. \citet{yang2024exploring} proposed to generate test cases, provide them along with the code to an LLM for translation, and use test feedback to iteratively repair the generated translation. Separately, there has been a long line of work on using transformer models for generating natural language descriptions of code \cite{ahmad2022summarize, gao2023code, gong2022source, gao2022m2ts, wu2021code, tang2021ast}. \citet{tang2023explain} and \citet{saha2024specification} both propose to generate natural language explanations as an intermediate step for program translation. There has been comparatively less research into generating static specifications. \citet{endres2023formalizing} transform natural language descriptions of functions into assertions and use them to detect incorrect code. Parsel \cite{zelikman2023parsel} first generates a program sketch consisting of function-level descriptions and test cases, and then generates code to complete the sketch. \citet{park2024loud} explore synthesizing formal static specifications for \textit{nondeterministic} programs.
\section{Threats to Validity}
% \res{Write Threats to Validity}
% \bi
% \item Our static specifications are in natural language and cannot be precisely validated against the source program. A self-consistent specification, while \textit{likely} to be correct, is not \textit{guaranteed} to be correct. 
% % Using formal specifications would solve this problem, but the reason why we opted not to use formal specifications is because: 1) C has no succinct way to describe assertions involving complex predicates like “for all”. In the few examples that we tried, the assertions are either trivial, or approach the complexity of the original function. 2) Our preconditions and postconditions are written in natural language, and LLMs are best optimized to understand natural language.

% \item \sysname{} introduces a computational overhead compared to single-step translation using LLMs. Even in the best case, where we find a self-consistent specification in the very first step, translating a program involves at least 2 additional LLM calls. Thus, there is a tradeoff between better performance (\sysname{}) and fewer LLM calls (vanilla translation). However, once specifications have been generated for a single source program, they can be used for multiple downstream tasks without any additional overhead.

% \item Our dataset for full-project translation is small, and our results in that setting may not be statistically significant. Nevertheless, we present it as a proof of concept that our approach can work in a full-project setting, and plan to explore this in future work.
% \ei

Our results have several threats to validity. 

\noindent
\textbf{Static Specs expressed in NL.} Our static specs are written in natural language, making it difficult to validate them precisely against the source program. While a self-consistent specification is a strong indication of correctness, it does not guarantee correctness. The absence of formal verification introduces potential ambiguities or misinterpretations. One way to address this issue would be to use formal specifications. However, we opted against this approach for two primary reasons:  
\be
    \item The C programming language lacks a succinct and expressive mechanism for specifying complex logical assertions, such as quantification over data structures (e.g., ``for all'' conditions). 
    \item Our preconditions and postconditions are written in NL, aligning with how LLMs are trained and optimized mostly on text corpora. 
\ee

\noindent
\textbf{Computational Overhead of \sysname{}}  
\sysname{} has computational overhead over a single-step translation using LLMs. Even in the best case, where we find a self-consistent specification in the very first step, translating a program involves at least 2 additional LLM calls. Thus, there is a tradeoff between better performance (\sysname{}) and fewer LLM calls (vanilla translation). However, once specifications have been generated for a single source program, they can be used for multiple downstream tasks without any additional overhead.

\noindent
\textbf{Dataset Size and Statistical Significance}  
Our dataset for full-project translation is relatively small, limiting the statistical significance of our results in this setting. While our findings suggest that our approach can be extended to full-project translation, they should be interpreted as preliminary evidence rather than definitive proof of scalability. Nonetheless, we present this as a proof of concept, demonstrating that our method is viable in a full-project setting. Future work will involve scaling up our dataset and conducting more extensive evaluations to assess generalizability.

\section{Conclusion}

In this paper, we have presented \sysname{}, an approach to 1) generate self-consistent multi-modal specifications, and 2) use these specifications to improve the program translation performance of large language models. We show relative performance gains of up to 46\% across multiple models and languages pairs. Our results indicate the possibility for further research into generating high-quality program specifications as a promising way to improve the performance of real-world LLMs on code translation. Future work could explore generating specifications in formal language or assert statements, which can then be automatically cross-verified against the source code. Another promising direction is to use these specifications for downstream tasks other than translation.

\bibliographystyle{ACM-Reference-Format}
\bibliography{references}
\clearpage
\appendix
\section{Additional Diagrams}
\begin{figure*}[h!]
    \centering
    \includegraphics[width=\textwidth]{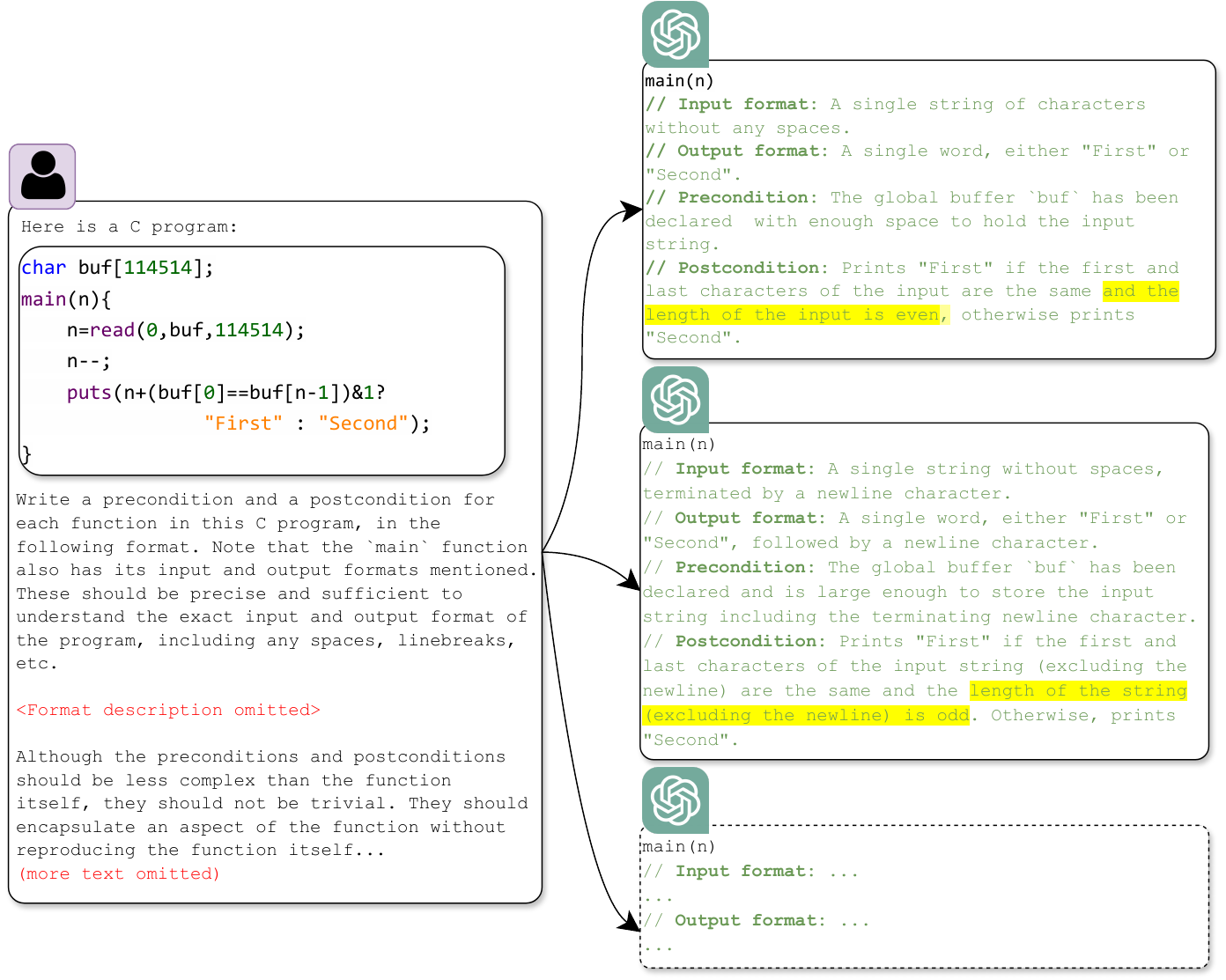}
    \caption{The first step of our approach involves generating multiple candidate specifications from a given program. This is the actual output of \texttt{gpt-4o} on our motivating example from \Cref{sec:motivating}.}
    \label{fig:forward-gen}
\end{figure*}

\begin{figure*}[h!]
    \centering
    \includegraphics[width=\textwidth]{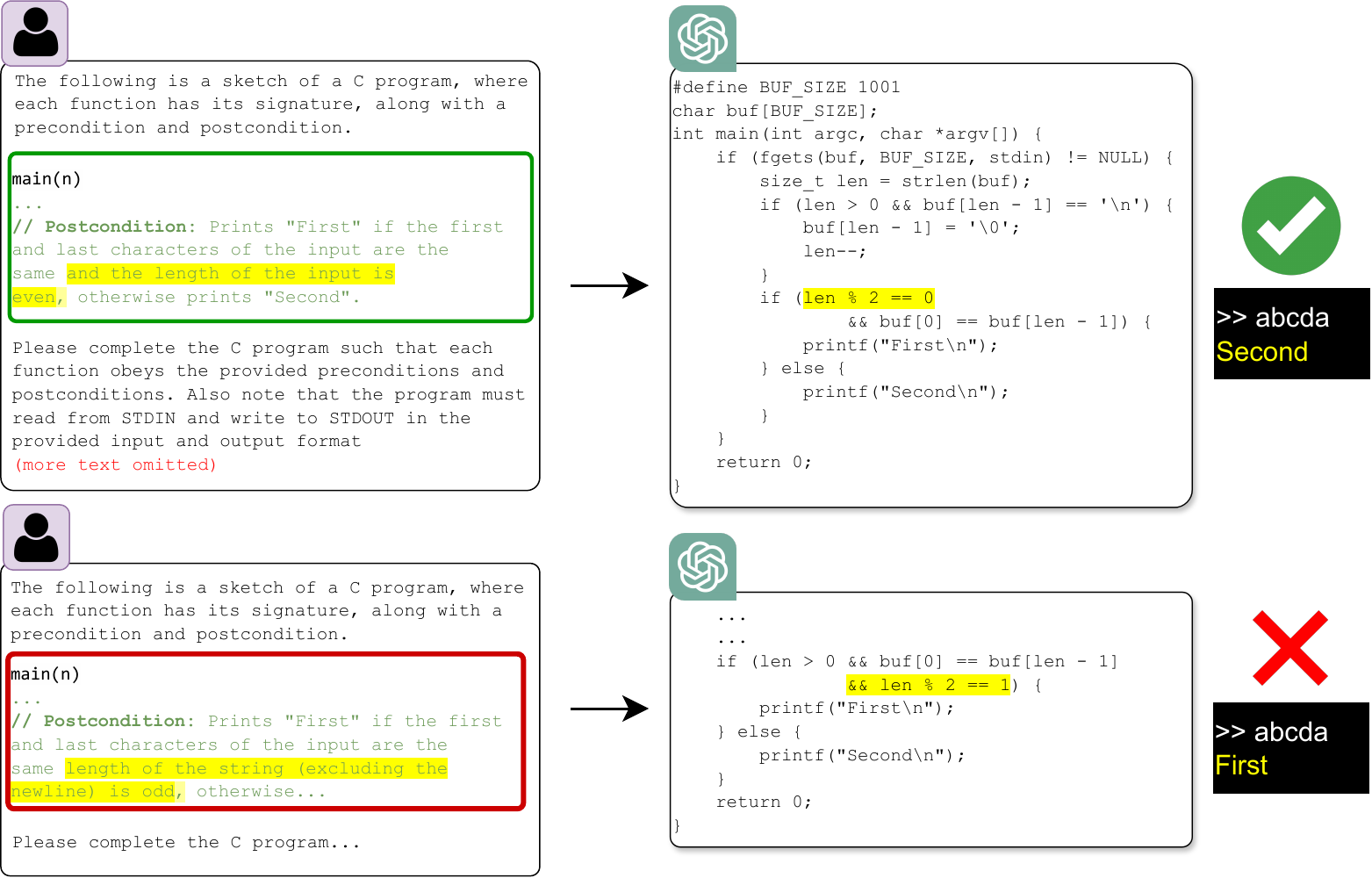}
    \caption{Filtering out incorrect specifications. We re-generate C code using each specification, and pick the specifications corresponding to re-generated C programs that pass the test case.}
    \label{fig:reverse-gen}
\end{figure*}

\begin{figure*}[h!]
    \centering
    \includegraphics[width=\textwidth]{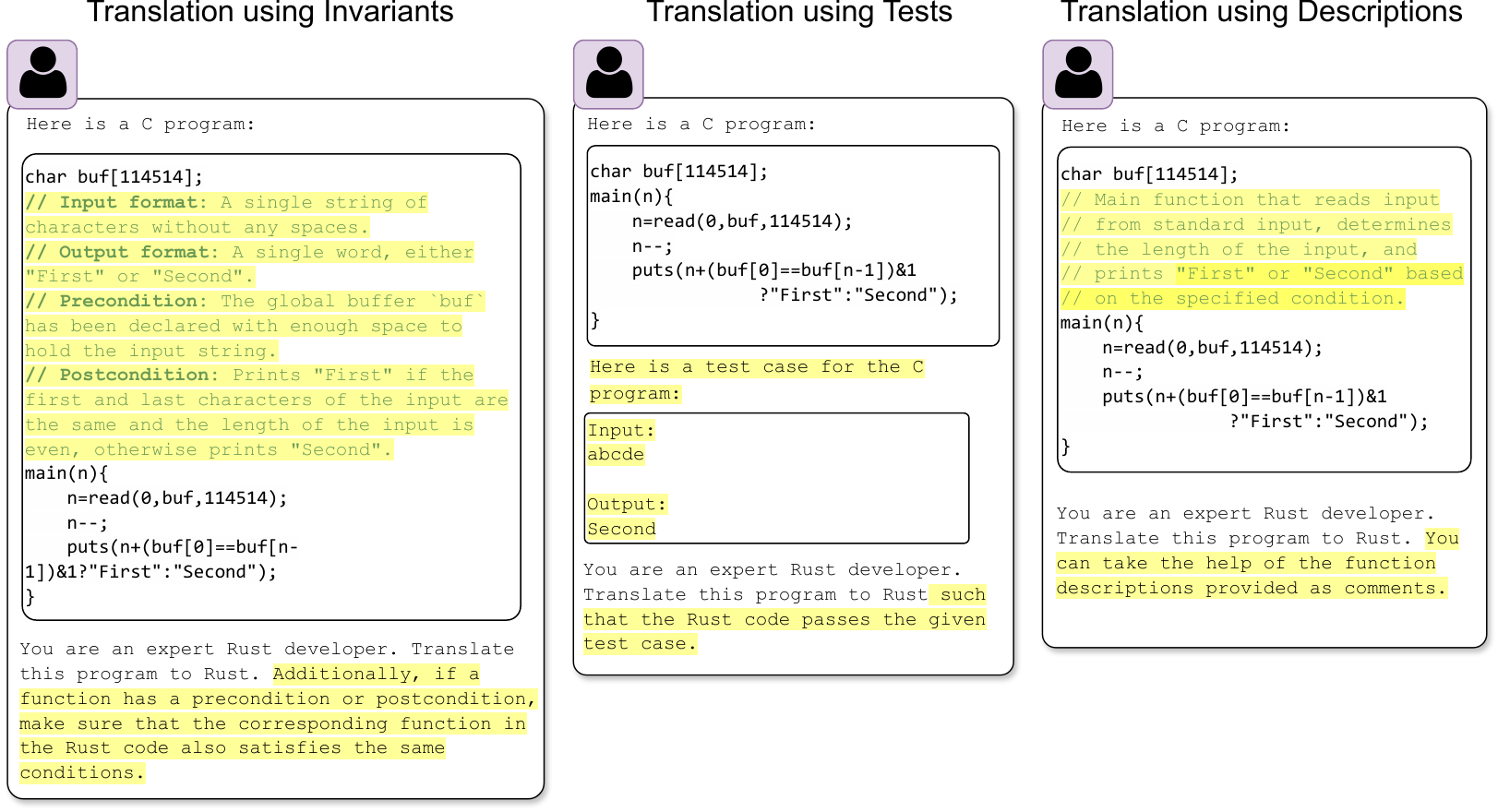}
    \caption{Using different modalities of specifications to perform specification-augmented translation. These are the actual prompts and specifications generated by \sysname{}.}
    \label{fig:modalities}
\end{figure*}

% \begin{figure*}[!h]
%     \centering
%     \includegraphics[width=0.7\textwidth]{figures/error_types.pdf}
%     \caption{Distribution of Error Types for \texttt{gpt-4o} C to Rust translation.}
%     \label{fig:error-types}
% \end{figure*}

\begin{figure*}[h!]
    \centering
    \begin{subfigure}[b]{0.9\columnwidth}
        \includegraphics[width=\textwidth]{figures/venn_rust.pdf}
        \caption{C to Rust}
    \end{subfigure}
    ~
    \begin{subfigure}[b]{0.45\columnwidth}
        \includegraphics[width=\textwidth]{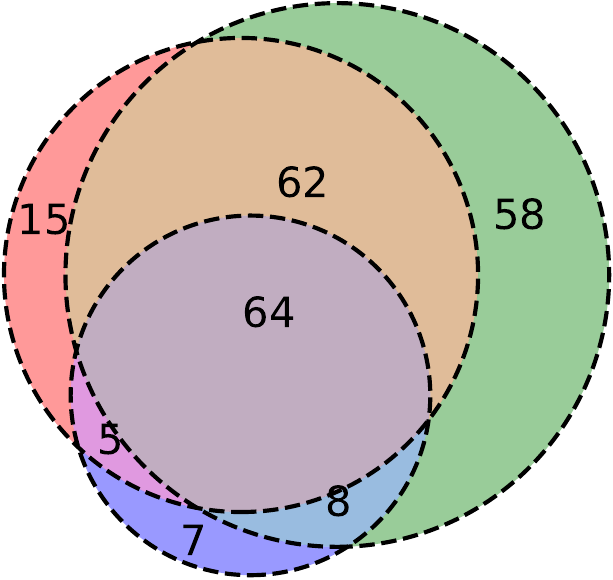}
        \caption{C to Go}
    \end{subfigure}
    \begin{subfigure}[b]{0.45\columnwidth}
        \includegraphics[width=0.95\textwidth]{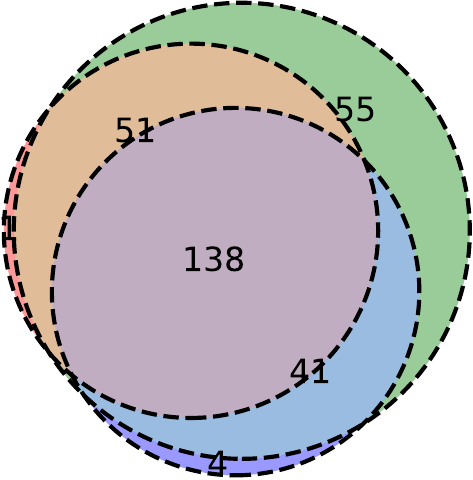}
        \caption{JavaScript to TypeScript}
    \end{subfigure}
    \caption{Our different specification modalities are complementary. This diagram shows how many out of the 300 translations are solved using each individual specification modality. All results are with \texttt{gpt-4o}.}
    \label{fig:venn_full}
\end{figure*}

\end{document}